\documentclass[iop,apj]{emulateapj}
\usepackage{apjfonts}
\usepackage{color}
\usepackage{graphicx}
\usepackage{hyperref}

\newcommand{\chisquare}{\ensuremath{\chi^{2}}}
\newcommand{\chisquaredata}{\ensuremath{\chi_{d}^{2}}}
\newcommand{\chisquaremodel}{\ensuremath{\chi_{m}^{2}}}
\newcommand{\sigmarn}{\ensuremath{\sigma_{\rm RN}}}
\newcommand{\thetavect}{\ensuremath{\vec{\theta}}}
\newcommand{\pmlr}{PMLR}

\newcommand{\imfit}{\textsc{imfit}}
\newcommand{\Imfit}{\textsc{Imfit}}
\newcommand{\makeimage}{\textsc{makeimage}}

\shorttitle{Imfit: Galaxy Image Fitting}
\shortauthors{Erwin}

\begin{document}

\title{Imfit: A Fast, Flexible New Program for Astronomical Image Fitting}

\author{Peter Erwin\altaffilmark{1,2}}

\altaffiltext{1}{Max-Planck-Insitut f\"{u}r extraterrestrische Physik,
Giessenbachstrasse, 85748 Garching, Germany}
\altaffiltext{2}{Universit\"{a}ts-Sternwarte M\"{u}nchen,
Scheinerstrasse 1, 81679 M\"{u}nchen, Germany}

\begin{abstract} 

I describe a new, open-source astronomical image-fitting program called
\imfit, specialized for galaxies but potentially useful for other
sources, which is fast, flexible, and highly extensible. A key
characteristic of the program is an object-oriented design which allows
new types of image components (2D surface-brightness functions) to be
easily written and added to the program. Image functions provided with
\imfit{} include the usual suspects for galaxy decompositions
(S{\'e}rsic, exponential, Gaussian), along with Core-S\'ersic and
broken-exponential profiles, elliptical rings, and three components
which perform line-of-sight integration through 3D luminosity-density
models of disks and rings seen at arbitrary inclinations.

Available minimization algorithms include Levenberg-Marquardt,
Nelder-Mead simplex, and Differential Evolution, allowing trade-offs
between speed and decreased sensitivity to local minima in the fit
landscape. Minimization can be done using the standard \chisquare{}
statistic (using either data or model values to estimate per-pixel
Gaussian errors, or else user-supplied error images) or Poisson-based
maximum-likelihood statistics; the latter approach is particularly
appropriate for cases of Poisson data in the low-count regime. I show
that fitting low-S/N galaxy images using \chisquare{} minimization and
individual-pixel Gaussian uncertainties can lead to significant biases
in fitted parameter values, which are avoided if a Poisson-based
statistic is used; this is true even when Gaussian read noise is
present. 

\end{abstract}


\section{Introduction} 

Galaxies are morphologically complex entities. Even seemingly simple
systems like elliptical galaxies can have outer envelopes and distinct
cores or nuclei,  while so-called ``bulge-less'' spiral galaxies can
still have nuclear star clusters and disks with complex radial or
vertical profiles. In order to accurately describe the structure of
galaxies, it is often necessary to decompose galaxies into component
substructures. Even single-component systems are often modeled with
analytic functions in order to derive quantitative measurements such as
scale lengths or half-light radii, S\'ersic indices, etc.

The traditional method for dealing with this complexity has been to
model 1D surface-brightness profiles of galaxies -- derived from 2D
images -- as the sum of separate, additive components (e.g., bulge +
disk); pioneering examples of this include work by
\citet{kormendy77b}, \citet{burstein79}, \citet{tsikoudi79,tsikoudi80},
\citet{boroson81}, \citet{send82}, and
\citet{hickson82}. While this 1D approach can be conceptually and
computationally simple, it has a number of limitations, above and beyond
the fact that it involves discarding most of the data contained in an
image. To begin with, there are uncertainties about \textit{what type}
of 1D profile to use -- should one use major-axis cuts or profiles from
ellipse fits to isophotes, should the independent variable be semi-major
axis or mean radius, etc.  It is also difficult to correctly account for
the effects of image resolution when fitting 1D profiles; attempts to do
so generally require simple analytic models of the point-spread function
(PSF), extensive numerical integrations, and the assumption of circular
symmetry for the PSF, the surface-brightness function, or both
\citep[e.g.,][]{pritchet81,saglia93,trujillo-moffat01,rusli13b}.
Furthermore, there are often intrinsic degeneracies involved: images of
galaxies with non-axisymmetric components such as bars can yield 1D
profiles resembling those from galaxies with axisymmetric bulges, which
makes for considerable ambiguity in interpretation. Finally, if one is
interested in the properties of non-axisymmetric components (bars,
elliptical rings, spiral arms) themselves, it is generally impossible to
extract these from 1D profiles.

A better approach in many cases is to directly fit the images with
2D surface-brightness models. Early approaches along this line
include those of \citet{capaccioli87}, \citet{shaw89}, and
\citet{scorza90}. The first general, self-consistent 2D bulge+disk
modeling of galaxy images -- that is, constructing a full 2D model
image, comparing its intensity values with the observed image pixel-by-pixel,
and iteratively updating the parameters until the \chisquare{} is
minimized -- was that of \citet{byun95}, with \citet{de-jong96} being
the first to include extra, non-axisymmetric components (bars) in
fitting galaxy images. An interesting alternate approach developed at
roughly the same time was the Multi-Gaussian Expansion method
\citep{monnet92,emsellem94,cappellari02}, which involves modeling both
PSF and image as the sum of an arbitrary number of elliptical Gaussians;
the drawback is the difficulty that lies in trying to associate sets of
Gaussians with particular structural components and parameters.

%





The most commonly used galaxy-fitting codes at the present time are
probably \textsc{gim2d}
\citep{simard98,simard02},\footnote{\href{https://www.astrosci.ca/users/
GIM2D/}{https://www.astrosci.ca/users/GIM2D/}} \textsc{galfit}
\citep{peng02,peng10},\footnote{\href{http://users.obs.carnegiescience.
edu/peng/work/galfit/galfit.html}{http://users.obs.carnegiescience.edu/
peng/work/galfit/galfit.html}} \textsc{budda}
\citep{desouza04,gadotti08},\footnote{\href{http://www.sc.eso.org/~
dgadotti/budda.html}{http://www.sc.eso.org/\~{}dgadotti/budda.html}} and
MGE
\citep{emsellem94,cappellari02}.\footnote{\href{http://www-astro.physics
.ox.ac.uk/~mxc/software/}{http://www-astro.physics.ox.ac.uk/\~{}mxc/
software/}} \textsc{gim2d} is specialized for bulge-disk decompositions
and is implemented as an \textsc{iraf} package, using the Metropolis
algorithm to minimize the total \chisquare{} for models containing an
exponential disk and a S\'ersic bulge. \textsc{budda} is written in
\textsc{fortran} and is also specialized for bulge-disk decompositions,
though it includes a wider variety of possible components: exponential
disk (with optional double-exponential profile), S\'ersic bulge,
S\'ersic bar, analytic edge-on disk, and nuclear point source. It uses a
version of the Nelder-Mead simplex method \citep{nelder-mead}, also
known as the ``downhill simplex'', for \chisquare{} minimization.
\textsc{galfit}, which is written in C, is the most general of these
codes, since it allows for arbitrary combinations of components
(including components with different centers, which allows the
simultaneous fitting of overlapping galaxies) and includes the largest
set of possible components; the latest version \citep{peng10} includes
options for spiral and other parametric modulation of the basic
components. \textsc{galfit} uses a version of the fast
Levenberg-Marquardt gradient-search method
\citep{levenberg44,marquardt63} for its \chisquare{} minimization. MGE,
available in IDL and Python versions, is rather different from the other
codes in that it uses what is effectively a non-parametric approach,
fitting images using the sum of an arbitrary number of elliptical
Gaussians (it is similar to \textsc{galfit} in using the
Levenberg-Marquardt method for \chisquare{} minimization during the
fitting process.)

For most astronomical image-fitting programs the source code is not
generally available, or else is encumbered by non--open-source licenses.
Even when the code \textit{is} available, it is not easy to extend the
built-in sets of predefined image components. The simplest codes provide
only elliptical components with exponential and S\'ersic surface
brightness profiles; more sophisticated codes such as \textsc{budda} and
(especially) \textsc{galfit} provide a larger set of components,
including some sophisticated ways of perturbing the components in the
case of \textsc{galfit}. But if one wants to add completely new
functions, this is not easy. (The case of MGE is somewhat different,
since it does not allow parametric functions at all.)

As an example of why one might want to do this, consider the case of
edge-on (or nearly edge-on) disk galaxies. Both \textsc{budda} and
\textsc{galfit} include versions of the analytical solution for a
perfectly edge-on, axisymmetric, radial-exponential disk of
\citet{vanderkruit81a}, with a ${\rm sech}^2$ function for the vertical
light distribution. But real galaxy disks are not always perfectly
edge-on, do not all have single-exponential radial structures, and their
vertical structure may in some cases be better described by a
sech or exponential profile, or something in between
\citep[e.g.,][]{vanderkruit88,degrijs97,pohlen04a,yoachim06}. Various
authors studying edge-on disks have suggested that models using radial
profiles other than a pure exponential would be best fit via
line-of-sight integration through 3D luminosity-density models
\citep[e.g.,][]{vanderkruit81a,pohlen00b,pohlen04a}. More sophisticated
approaches could even involve line-of-sight integrations that account for
scattering and absorption by dust
\citep[e.g.,][]{xilouris97,xilouris98,xilouris99}.

Another potential disadvantage of existing codes is that they rely on
the Gaussian approximation of Poisson statistics for the fitting
process. While this is eminently sensible for dealing with many CCD and
near-IR images, it can in some cases produce biases when applied
to images with low count rates (see \citealt{humphrey09} and
Section~\ref{sec:biases} of this paper). This is why packages for
fitting X-ray data, such as \textsc{sherpa} \citep{sherpa}, often include
alternate statistics for fits.

In this paper, I present \imfit{}, a new, open-source image-fitting code
designed to overcome some of the limitations mentioned above. In
particular, \imfit{} uses an object-oriented design which makes it
relatively easy to add new, user-designed image components; it also
provides multiple fitting algorithms and statistical approaches. It can
also be extremely fast, since it is able to take advantage of multiple CPU
cores on the same machine to execute calculations in parallel.

The outline of this paper is as follows. Section~\ref{sec:gen-outline}
provides a quick sketch of how the program works, while
Section~\ref{sec:modelimage} details the process of generating model
images and the configuration files which describe the models. The
different underlying statistical models and minimization algorithms used
in the fitting process are covered in Section~\ref{sec:fitting}; methods
for estimating confidence intervals for fitted parameters are discussed
in Section~\ref{sec:confidence-intervals}. The default 2D image
functions which can be used in models are presented in
Section~\ref{sec:image-funcs}; this includes functions which perform
line-of-sight integration through 3D luminosity-density models
(Section~\ref{sec:image-funcs-3d}). After a brief discussion of coding
details (Section~\ref{sec:programming}), two examples of using \imfit{}
to model galaxy images are presented in Section~\ref{sec:examples}: the
first involves fitting a moderately-inclined spiral galaxy with disk,
bar, and ring components, while the second fits an edge-on spiral galaxy
with thin and thick edge-on disk components. Finally,
Section~\ref{sec:biases} discusses possible biases to fitted parameters
when the standard \chisquare{} statistic is used in the presence of
low-count images, using both model images and real images of elliptical
galaxies. An Appendix discusses the relative sizes and accuracies
of parameter error estimates using the two methods available in \imfit.

To avoid any confusion, I note that the program described in this
paper is unrelated to tasks with the same name and somewhat similar (if
limited) functionality in pre-existing astronomical software, such as
the ``imfit'' tasks in the radio-astronomy packages \textsc{aips} and
\textsc{miriad} and the ``images.imfit''  package in \textsc{iraf}.

\section{General Outline of the Program}\label{sec:gen-outline} 

\Imfit{} begins by processing command-line options and then reads in the
data image, along with any optional, user-specified PSF, noise, and mask
images (all in \textsc{fits} format). The configuration file is also
read; this specifies the model which will be fit to the data image,
including initial parameter values and parameter limits, if any (see
Section~\ref{sec:config-file}).

The program then creates an instance of the ModelObject class, which
holds the relevant data structures, instances of the image functions
specified by the configuration file, and the general code necessary for
computing a model image.  If \chisquare{} minimization (the default) is
being done, a noise image is constructed, either from a user-specified
\textsc{fits} file already read in or by internally generating one,
assuming the Gaussian approximation for Poisson noise. The noise image
is then converted to $1/\sigma^2$ form and combined with the mask image,
if any, to form a final weight image used for calculating the
\chisquare{} value. (If model-based \chisquare{} minimization has been
specified, then the noise image, which is based on the model image, is
recalculated and combined with the mask image every time a new model
image is computed; if a Poisson maximum-likelihood statistic ($C$
or \pmlr; see Section~\ref{sec:poisson}) is being used for
minimization, then no noise image is read or created and the weight
image is constructed directly from the mask image. See
Section~\ref{sec:statistics} for more on the different statistical
approaches.)

The actual fitting process is overseen by one of three possible
nonlinear minimization algorithms, as specified by the user. These
algorithms proceed by generating or modifying a set of parameter values
and feeding these values to the aforementioned model object, which in
turn calculates the corresponding model image, convolves it with the PSF
(if PSF convolution is part of the model), and then calculates the fit
statistic (e.g., \chisquare) by comparing the model image with the
stored data image. The resulting fit statistic is returned to the
minimization algorithm, which then updates the parameter values and
repeats the process according to the details of the particular method,
until the necessary stop criterion is reached -- e.g., no further
significant reduction in the fit statistic, or a maximum number of
iterations. Finally, a summary of the fit results is printed to the
screen and saved to a file, along with any additional user-requested
outputs (final model image, final residual image, etc.).

\section{Constructing the Model Image}\label{sec:modelimage} 

\subsection{Configuration File}\label{sec:config-file} 

\begin{figure*}
\begin{center}
\includegraphics[scale=0.7]{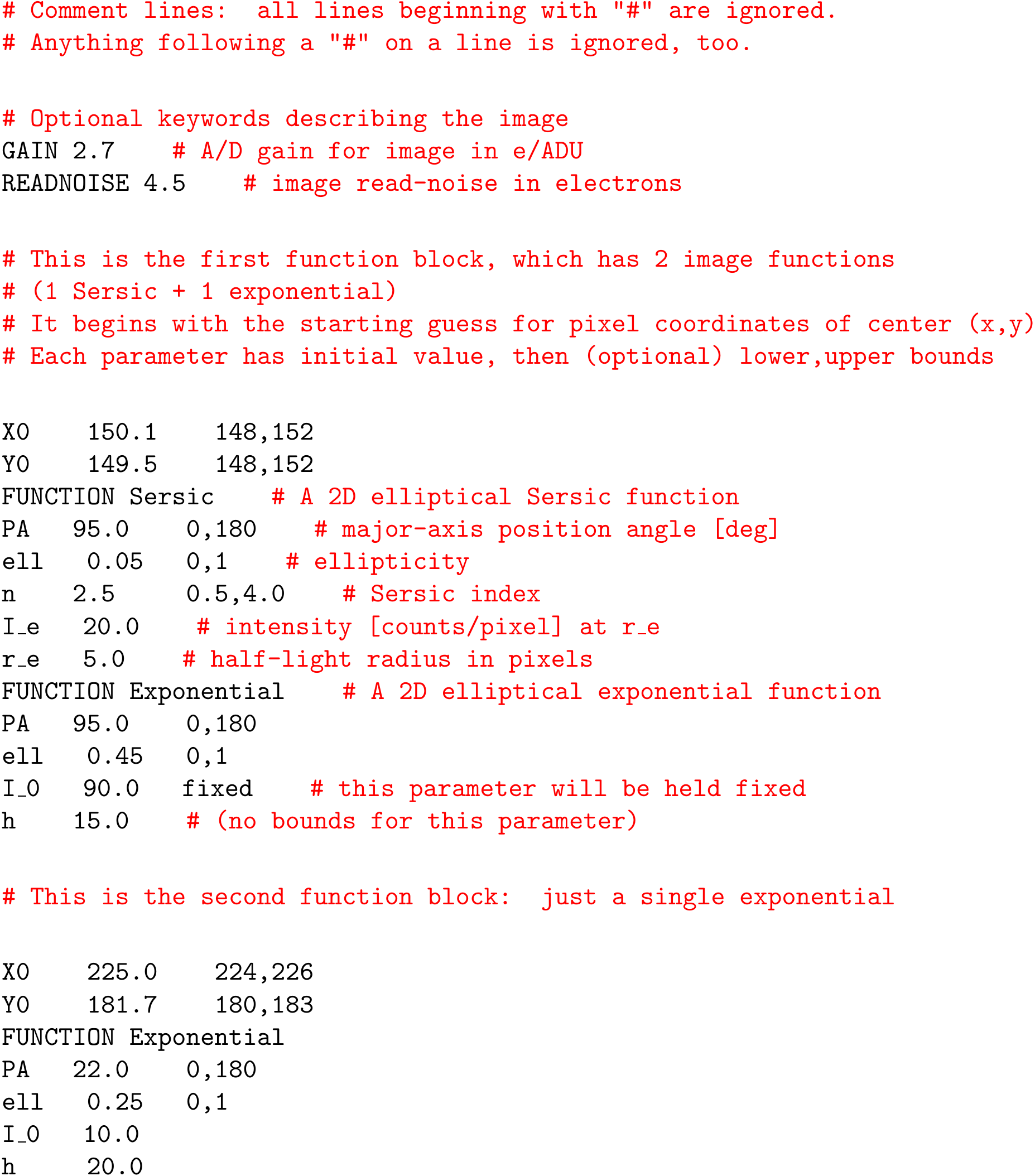}
\end{center}

\caption{Example of a configuration file for \imfit. Comments are colored
red.\label{fig:config-file}}

\end{figure*}

The model which will be fit to the data image is specified by a
configuration file, which is a text file with a relatively
simple and easy-to-read format; see Figure~\ref{fig:config-file} for
an example.

The basic format for this file is a set of one or more ``function
blocks'', each of which contains a shared center (pixel coordinates) and
one or more image functions.  A function block can, for
example, represent a single galaxy or other astronomical object, which
itself has several individual components (e.g., bulge, disk, bar, ring, nucleus,
etc.) specified by the individual image functions. Thus, for a basic
bulge/disk decomposition the user could create a function block
consisting of a single S{\'e}rsic function and a single Exponential
function. There is, however, no a priori association of any particular
image function or functions with any particular galaxy component, nor is
there any requirement that a single object must consist of only one
function block. The final model is the sum of the contributions from all
the individual functions in the configuration file. The number of image
functions per function block is unlimited, and the number of function
blocks per model is also unlimited.

Each image function is listed by name (e.g., ``\texttt{FUNCTION
Sersic}''), followed by the list of its parameters. For each parameter,
the user supplies an initial guess for the value, and (optionally)
either a comma-separated, two-element list of lower and upper bounds for
that parameter or the keyword ``fixed'' (indicating that the parameter
will remain constant during the fit).\footnote{Parameter bounds can be
used with any of the minimization algorithms; with
the Differential Evolution algorithm, they are actually
\textit{required}, though in that case the initial values are ignored;
see Section~\ref{sec:minimization}.}

The total set of all individual image-function parameters, along with
the central coordinates for each function block, constitutes the
parameter vector for the minimization process.

\subsection{Image Functions} 

An image function can be thought of as a black box which accepts a set
of parameter values for its general setup, and then accepts individual
pixel coordinates $(x,y)$ and returns a corresponding computed intensity
(i.e., surface brightness) value for that pixel. The total intensity for a
given pixel in the model image (prior to any PSF convolution) is the sum
of the individual values from each image function.

This design means that the main program needs to know nothing about
the individual image functions except the number of parameters they
take, and which subset of the total parameter vector corresponds to a
given image function.  The actual calculations carried out by an image function
can be as simple or as complex as the user requires, ranging from returning
a constant value for each pixel (e.g., the FlatSky function) to performing
line-of-sight integration through a 3D luminosity density model (e.g., the
ExponentialDisk3D function); user-written image functions could even
perform modest simulations in the setup stage.\footnote{One should bear
in mind that even relatively simple fits will typically require dozens
to hundreds of function evaluations during the minimization process, so
complex simulations will mean a lengthy fitting process.}

The list of currently available image functions, along with descriptions
for each, is given in Section~\ref{sec:image-funcs}.

\subsection{PSF Convolution} 

To simulate the effects of atmospheric seeing and telescope optics,
model images can be convolved with a PSF image. The latter can be any
\textsc{fits} file which contains the point spread function. PSF
images should ideally be square with sides measuring an odd number of pixels,
with the peak of the PSF centered in the central pixel of the image.
(Off-center PSFs can be used, but the resulting convolved model images
will of course be shifted.) \Imfit{} automatically normalizes the PSF
image when it is read in.

The actual convolution follows the standard approach of using Fast
Fourier Transforms of the internally-generated model image and the PSF
image, multiplied together, with the output convolved model image being
the inverse transform of the product image. The transforms are done with
the FFTW library \citep[``Fastest Fourier Transform in the
West'',][]{fftw},\footnote{\href{http://www.fftw.org}{http://www.fftw.org}} 
which has the advantage of being able to perform transforms on
images of arbitrary size (i.e., not just images with power-of-two
sizes); in addition, it is well-tested and fast, and can use multiple
threads to take advantage of multiple processor cores.

To avoid possible edge effects in the convolution, the internal
model-image array is expanded on all four sides by the width and height
of the PSF image, and all calculations prior to the convolution phase
use this full (expanded) image. (For example, given a $1000 \times
1000$-pixel data image and a $15 \times 15$-pixel PSF image, the
internal model image would be $1030 \times 1030$ pixels in size.) This ensures
that model pixels corresponding to the edge of the data image are the
result of convolution with an extension of the model, rather than with
zero-valued pixels or the opposite side of the model image. This is in
\textit{addition} to the zero-padding applied to the top and right-hand
sides of the model image during the convolution phase. (I.e., the
example $1030 \times 1030$-pixel expanded model image would be
zero-padded to $1045 \times 1045$ pixels before computing its Fourier
transform, to match with the zero-padded PSF image of the same size.)

\subsection{Makeimage: Generating Model Images Without Fitting} 

A companion program called \makeimage{} is included in the \imfit{}
package, built from the same codebase as \imfit{} itself. This program
implements the complete model-image construction process, including PSF
convolution, and then simply saves the resulting model image as a \textsc{fits}
file. It can optionally save separate images, one for each of the
individual image functions that make up the model. Since it uses the
same configuration-file format as \imfit, it can use the output best-fit
parameter file that \imfit{} produces (or even an input \imfit{}
configuration file). 

It also has an optional mode which estimates the fractional flux
contributions of the individual components in the model, by summing up
the total flux of the individual components on a pixel-by-pixel basis
using a very large internal image (by default, $5000 \times 5000$
pixels). Although analytic expressions for total flux exist for some
common components, this is not true for all components -- and one of the
goals of \imfit{} is to allow users to create and use new image
functions without worrying about whether they have simple analytic
expressions for the total flux. This mode can be used to help determine
such things as bulge/total and other ratios after a fit is found,
although it is up to the user to decide which of the components is the
``bulge'', which is the ``disk'', and so forth.

\section{The Fitting Process}\label{sec:fitting} 

\subsection{The Statistical Background and Options}\label{sec:statistics} 

Given a vector of parameter values \thetavect, a model
image is generated with per-pixel predicted data values $m_{i}$, which are
then compared with the \textit{observed} per-pixel data values $d_{i}$. The
goal is to find the \thetavect{} which produces the best match between
$m_{i}$ and $d_{i}$, subject to the constraints of the underlying
statistical model.

The usual approach is based on the maximum-likelihood principle (which
can be derived from a Bayesian perspective if, e.g., one assumes
constant priors for the parameter values), and is conventionally known as 
maximum-likelihood estimation \citep[MLE; e.g.,][]{pawitan01}. To start, one considers
the per-pixel likelihood $p_{i}(d_{i} | m_{i})$, which is the
probability of observing $d_{i}$ given the model prediction $m_{i}$ and
the underlying statistical model for how the data are generated.

The goal then becomes finding the set of model parameters which maximizes the
total likelihood $\mathcal{L}$, which is simply the product over all $N$ pixels
of the individual per-pixel likelihoods:
\begin{equation}
\mathcal{L} \; = \; \prod_{i=1}^{N} \, p_{i}.
\end{equation}
It is often easier to work with the logarithm of the total likelihood,
since this converts a product over pixels into a sum over pixels, and
can also simplify the individual per-pixel terms. As most nonlinear
optimization algorithms are designed to \textit{minimize} their
objective function, one can use the \textit{negative} of the
log-likelihood. Thus, the goal of the fitting process becomes
minimization of the following:
\begin{equation}
-\ln \mathcal{L} \; = \; -\sum_{i=1}^{N} \, \ln p_{i}.
\end{equation}
During the actual minimization process, this can often be further
simplified by dropping any additive terms in $\ln p_{i}$ which do not
depend on the model, since these are unaffected by changes in the model
parameters and are thus irrelevant to the minimization.

In some circumstances, multiplying the negative log-likelihood by 2 produces
a value which has the property of being distributed like the
\chisquare{} distribution \citep[e.g.,][and references therein]{cash79};
thus, it is conventional to treat $-2 \ln \mathcal{L}$ as the statistic to
be minimized.

\subsubsection{The (Impractical) General Case: Poisson + Gaussian  Statistics} 

The data in astronomical images typically consist of detections of
individual photons from the sky + telescope system (including
photons from the source, the sky background, and possibly thermal
backgrounds in the telescope) in individual pixels, combined with
possible sources of noise due to readout electronics, digitization, etc.

Photon-counting statistics obey the Poisson distribution, where the probability
of detecting $x$ photons per integration, given a true rate of $m$, is
\begin{equation}
P(x) \; = \; \frac{m^{x} e^{-m}}{x!}.
\end{equation}\label{eq:poisson}

Additional sources of (additive) noise such as read noise tend to follow
Gaussian statistics with a mean of 0 and a dispersion of $\sigma$, so
that the probability of measuring $d$ counts after the readout process,
given an input of $x$ counts from the Poisson process, is
\begin{equation}\label{eq:gaussian}
P(d) \;  = \; \frac{1}{\sqrt{2 \pi} \sigma} \exp\left[ 
\frac{-(d - x)^{2}}{2 \sigma^{2}} \right].
\end{equation}

The general case for most astronomical images thus involves both Poisson
statistics (for photon counts) and Gaussian statistics (for read noise
and other sources of additive noise). Unfortunately, even though the individual
elements are quite simple, the combination of a Gaussian process acting on
the output of a Poisson process leads to the following rather frightening per-pixel likelihood 
\citep[e.g.,][]{llacer91,nunez93}:
\begin{equation} 
p_{i}(d_{i} | m_{i}) \; = \; \sum_{x_{i}=0}^{\infty} \frac{m_{i}^{x_{i}} e^{-m_{i}}}{x_{i}!} 
\frac{1}{\sqrt{2 \pi} \sigma} \exp \left( \frac{-(d_{i} - x_{i})^{2}}{2 \sigma^{2}} \right).
\end{equation}
The resulting negative log-likelihood for the total image (dropping terms which do not
depend on the model) is
\begin{equation} 
-\ln \mathcal{L} \; = \; \sum_{i=1}^{N} \left( m_{i} \, - \, \ln \left[ \sum_{x_{i}=0}^{\infty} 
\frac{m_{i}^{x_{i}}}{x_{i}!} \exp \left( \frac{-(d_{i} - x_{i})^{2}}{2 \sigma^{2}} \right) \right] \right).
\end{equation}
Since this still contains an infinite series of exponential and
factorial terms, it is clearly rather impractical for fitting images
rapidly.

\subsubsection{The Simple Default: Pure Gaussian Statistics} 

Fortunately, there is a way out which is often (though not always) appropriate 
astronomical images. This is to use the fact that the Poisson
distribution approaches a Gaussian distribution when the counts become
large. In this approximation the Poisson distribution is replaced by
a Gaussian with $\sigma = \sqrt{m}$. It is customary to assume
this is valid when the counts are $\ga 20$ per pixel
\citep[e.g.,][]{cash79}, though \citet{humphrey09} point out that biases
in the fitted parameters can be present even when counts are higher than
this; see Section~\ref{sec:biases} for examples in the case of 2D fits. 

Since the contribution from read noise is also nominally Gaussian, the two can be
added in quadrature, so that the per-pixel likelihood function is just
\begin{equation}
p_{i}(d_{i} | m_{i}) \; = \; \frac{1}{\sqrt{2 \pi} \sigma_{i}} \exp \left[ 
\frac{-(d_{i} - m_{i})^{2}}{2 \sigma_{i}^{2}}\right],
\end{equation}
where $\sigma_{i}^{2} = \sigma_{m_{i}}^{2} + \, \sigmarn^{2} = m_{i} + \, 
\sigmarn^{2}$, with \sigmarn{} being the dispersion of the read-noise term. Twice the negative
log-likelihood of the total problem then becomes (dropping terms which do not
depend on the model) the familiar \chisquare{} sum:
\begin{equation}\label{eqn:chi2}\label{eq:chi2}
-2 \ln \mathcal{L} \; = \; \chi^{2} \; = \; \sum_{i = 1}^{N} \frac{(d_{i} - m_{i})^{2}}{\sigma_{i}^{2}}.
\end{equation}

This is the default approach used by \imfit: minimizing the \chisquare{}
as defined in Eqn.~\ref{eqn:chi2}. 

The approximation of the Poisson contribution to $\sigma_{i}$ is based
on the model intensity $m_{i}$. Traditionally, it is quite common to estimate
this from the \textit{data} instead, so that $\sigma_{i}^{2} =
\sigma_{d_{i}}^{2} + \, \sigmarn^{2} = d_{i} + \, \sigmarn^{2}$.  This
has the nominal advantage of only needing to be calculated once, at the
start of the minimization process, rather than having to be recalculated
every time the model is updated.\footnote{In practice, the time spent by
\imfit{} is dominated by the per-pixel model calculations, so any extra
time spent re-estimating the per-pixel $\sigma_{i}$ values is often
negligible.} However, the bias resulting from using data-based errors in
the low-count regime can be worse than the bias introduced by using
model-based $\sigma_{i}$ values (see Section~\ref{sec:biases}). Both
approaches are available in \imfit, with data-based $\sigma_{i}$
estimation being the default. The data-based and model-based approaches
are often referred to as ``Neyman's \chisquare{}'' and ``Pearson's
\chisquare{}'', respectively; in this paper I use the symbols \chisquaredata{}
and \chisquaremodel{} to distinguish between them. 

In the case of ``error'' images generated by a data-processing pipeline,
the corresponding $\sigma_{i}$ or $\sigma_{i}^{2}$ (variance) values can
easily be used in Equation~\ref{eq:chi2} directly, under the assumption
that the final per-pixel error distributions are still Gaussian.


\subsubsection{The Simple Alternative: Pure Poisson Statistics}\label{sec:poisson} 

So why not always use the Gaussian \chisquare{} approximation, as is done in most
image-fitting packages?

In the absence of any noise terms except Poisson statistics -- something often true of
high-energy detectors, such as X-ray imagers -- the individual-pixel likelihoods
are just the probabilities of a Poisson process with mean $m_{i}$, where the probability
of recording $d_{i}$ counts is
\begin{equation}\label{eq:poisson-likelihood}
p_{i}(d_{i} | m_{i}) \; = \; \frac{m_{i}^{d_{i}} e^{-m_{i}}}{d_{i}!}
\end{equation}
This leads to a very simple version of the negative log-likelihood, often
referred to as the ``Cash statistic'' $C$, after its derivation in \citet{cash79}:
\begin{equation}\label{eq:cashstat}
-2 \ln \mathcal{L} \; = \; C \; = \; 2 \sum_{i = 1}^{N} \left( m_{i} - d_{i} \ln m_{i} \right),
\end{equation}
where the factorial term has been dropped because it does not depend on the model.

A useful alternative is to construct a statistic from the likelihood
ratio test -- that is, a maximum likelihood ratio (MLR) statistic -- which is
the ratio of the likelihood to the maximum possible likelihood for a given
dataset. In the case of Poisson likelihood, the latter is the likelihood when the
model values are exactly equal to the data values $m_{i} = d_{i}$ \citep[e.g.,][]{hauschild01}, 
and so the likelihood ratio $\lambda$ is
\begin{equation}
\lambda \; = \; \mathcal{L}/\mathcal{L}_{\mathrm{max}} \; = \; \prod_{i=1}^{N} \frac{m_{i}^{d_{i}} e^{-m_{i}}}{d_{i}^{d_{i}} e^{-d_{i}}}
\end{equation}
and the negative log-likelihood version (henceforth \pmlr) is
\begin{equation}\label{eq:pmlr}
\mathrm{\pmlr} \; = \; -2 \ln \lambda \; = \; 2 \sum_{i = 1}^{N} \left( m_{i} - d_{i} \ln m_{i} + d_{i} \ln d_{i} - d_{i} \right).
\end{equation}
(This is the same as the ``CSTAT'' statistic available in the \textsc{sherpa}
X-ray analysis package and the ``Poisson likelihood ratio'' described by
\citealt{dolphin02}.) Comparison with Equation~\ref{eq:cashstat} shows that \pmlr{} is
identical to $C$ apart from terms which depend on the data only and
thus do not affect the minimization. In the remainder of this paper, I will
refer to $C$ and \pmlr{} collectively as \textit{Poisson MLE} statistics.

Since minimizing \pmlr{} will produce the same best-fitting
parameters as minimizing $C$, one might very well wonder what is the
point in introducing \pmlr. There are two practical advantages in using
it. The first is that in the limit of large $N$, $-2 \ln \lambda$
statistics such as \pmlr{} approach a \chisquare{} distribution and can
thus be used as goodness-of-fit indicators \citep{wilks38,wald43}. The
second is that they are always $\ge 0$ (since $\lambda$ itself is by
construction always $\le 1$); this means they can be used with fast
least-squares minimization algorithms. This is the practical drawback to
minimizing $C$: unlike \pmlr, it can often have negative values, and
thus requires one of the slower minimization algorithms.

\citet{humphrey09} point out that using a Poisson MLE
statistic (e.g., $C$) is preferable to using \chisquaredata{} or
\chisquaremodel{} even when the counts are above the nominal limit of
$\sim 20$ per pixel, since fitting pure-Poisson data using the
\chisquaredata{} or \chisquaremodel{} Gaussian approximations can
lead to biases in the derived model parameters. Section~\ref{sec:biases}
presents some examples of this effect using both artificial and real
galaxy images, and shows that the effect persists even when moderate
(Gaussian) read noise is \textit{also} present.

Using a Poisson MLE statistic such as $C$ or \pmlr{} is also
appropriate when fitting simulated images, such as those made from
projections of $N$-body models, as long as the units are particles per
pixel or something similar.

For convenience, Table~\ref{tab:terminology} summarizes the main symbols and terms from
this section which are used elsewhere in the paper.


\begin{deluxetable}{ll}
\tablecaption{Terminology for Fits and Minimization\label{tab:terminology}}
\tablecolumns{2}
\tablehead{\colhead{Term} & \colhead{Explanation}}
\startdata
Poisson MLE & Maximum-likelihood estimation based on Poisson statistics \\
            & (includes both $C$ and \pmlr) \\
$C$ & Poisson MLE statistic from \citet{cash79} \\
\pmlr & Poisson MLE statistic from maximum likelihood ratio \\
\chisquaredata & Gaussian MLE statistic using data pixel values for $\sigma$ \\
               & (``Neyman's \chisquare{}'') \\
\chisquaremodel & Gaussian MLE statistic using model pixel values for $\sigma$ \\
                & (``Pearson's \chisquare{}'') \\
\enddata
\end{deluxetable}

%
%

\subsection{Implementation: Specifying Per-Pixel Errors and Masking} 

\Imfit{}'s default behavior, as mentioned above, is to use \chisquare{}
as the statistic for minimization. To do so, the individual, per-pixel
Gaussian errors $\sigma_{i}$ must be available. If a separate error or noise map
is not supplied by the user (see below), \imfit{} estimates $\sigma_{i}$ values from
either the data values or the model values, using the Gaussian
approximation to Poisson statistics. To ensure this estimate is as
accurate as possible, the data or model values $I_{i}$ must at some point be
converted from counts to actual detected photons (e.g., photoelectrons),
and any previously subtracted background must be accounted for.

By default, \imfit{} estimates the $\sigma_{i}$ values from the data
image by including the effects of A/D gain, prior subtraction of a (constant)
background, and read noise. Rather than converting the image
to electrons pixel$^{-1}$ and then estimating the $\sigma$ values,
\imfit{} generates $\sigma$ values in the same units as the input image:
\begin{equation}
\sigma^{2}_{I,i} \; = \; (I_{d, i} \, + \, I_{\mathrm{sky}})/g_{\mathrm{eff}} \: + \: N_{\mathrm{c}} \, \sigma_{\mathrm{RN}}^{2}/g_{\mathrm{eff}}^{2} \, ,
\end{equation}\label{eqn:error-est}
where $I_{d, i}$ is the data intensity in counts pixel$^{-1}$,
$I_{\mathrm{sky}}$ is any pre-subtracted sky background in the same
units, $\sigma_{\mathrm{RN}}$ is the read noise in electrons,
$N_{\mathrm{c}}$ is the number of separate images combined (averaged or
median) to form the data image, and $g_{\mathrm{eff}}$ is the
``effective gain'' (the product of the $A/D$ gain,  $N_{\mathrm{c}}$,
and optionally the exposure time if the image pixel values are actually
in units of counts s$^{-1}$ pixel$^{-1}$ rather than integrated counts
pixel$^{-1}$). If model-based \chisquare{} minimization is used,
then model intensity values $I_{m, i}$ are used in place of $I_{d, i}$
in Equation~\ref{eqn:error-est}. In this case, the $\sigma_{I,i}$ values
must be recomputed each time a new model image is generated, though in
practice this adds very little time to the overall fitting process.

If a mask image has been supplied, it is converted internally so that
its pixels have values $z_{i} = 1$ for valid pixels and $z_{i} = 0$
for bad pixels. Then the mask values are
divided by the variances to form a weight-map image, where individual pixels
have values of $w_{i} = z_{i} / \sigma_{I,i}^{2}$.  These weights are then used
for the actual \chisquare{} calculation:
\begin{equation}
\chisquare \; = \; \sum_{i = 1}^{N} w_{i} \, (I_{d, i} \, - \, I_{m, i})^{2}.
\end{equation}

Instead of data-based or model-based errors, the user can also supply an
error or noise map in the form of a \textsc{fits} image, such as might be
produced by a reduction pipeline. The individual pixel values in this image
can be Gaussian errors, variances ($\sigma^{2}$), or even
pre-computed weight values $w_{i}$.

In the case of Cash-statistic minimization, the sum $C$ is computed
directly based on Equation~\ref{eq:cashstat}; for \pmlr{} minimization,
Equation~\ref{eq:pmlr} is used. The ``weight map'' in either case is then
based directly on the mask image, if any (so all pixels in the resulting
weight map have values of $z_{i} = 0$ or 1). The actual minimized
quantities are thus
\begin{equation}
C \; = \; 2 \sum_{i = 1}^{N} z_{i} \left( m_{i} \, - \, d_{i} \ln m_{i} \right)
\end{equation}
and
\begin{equation}
{\rm \pmlr} \; = \; 2 \sum_{i = 1}^{N} z_{i} \left( m_{i} \, - \, d_{i} \ln m_{i} 
   + d_{i} \ln d_{i} - d_{i} \right)
\end{equation}
with $m_{i} = g_{\mathrm{eff}} (I_{m, i} + I_{\mathrm{sky}})$ and 
$d_{i} = g_{\mathrm{eff}} (I_{d, i} + I_{\mathrm{sky}})$.


\subsection{Minimization Algorithms}\label{sec:minimization} 

\subsubsection{Levenberg-Marquardt} 

The default minimization algorithm used by \imfit{} is a robust
implementation of the Levenberg-Marquardt (L-M) gradient search method
\citep{marquardt63}, based on the MINPACK-1 version of \citet{more78}
and modified by Craig Markwardt
\citep{markwardt09},\footnote{\href{http://purl.com/net/mpfit}{http://
purl.com/net/mpfit}} which includes optional lower and upper bounds on
parameter values. This version of the basic L-M algorithm also includes
auxiliary code for doing numerical differentiation of the objective
function, and thus the various image functions do not need to provide
their own derivatives, which considerably simplifies things when it
comes to writing new functions.

The L-M algorithm has the key advantage of being very fast, which is a
useful quality when one is fitting large images with a complex set
of functions and PSF convolution. It has the minor disadvantage of
requiring an initial starting guess for the parameter values, and it has
two more significant disadvantages. The first is that like
gradient-search methods in general it is prone to becoming trapped in
local minima in the objective-function landscape. The second is that it
is designed to work with least-squares objective functions, where the
objective function values are assumed to be always $\ge 0$. In fact, the
L-M algorithm makes use of a vector of the individual contributions from
each pixel to the total \chisquare, and these values as well (not just
the sum) must be nonnegative. For the \chisquare{} case, this is always
true; but this is \textit{not} guaranteed to be true for the Cash
statistic $C$. Thus, it would be quite possible for the L-M minimizer to fail to
find the best-fitting solution for a particular image, simply because
the solution has a $C$ value $< 0$. (Fortunately, minimizing
\pmlr{} leads to the same solution as minimizing $C$, and the individual
terms of \pmlr{} are always nonnegative.)

\subsubsection{Nelder-Mead Simplex} 

A second, more general algorithm available in \imfit{} is the
Nelder-Mead simplex method \citep{nelder-mead}, with constraints as
suggested by \citet{box65}, implemented in the NLopt
library.\footnote{Steven G. Johnson, The NLopt nonlinear-optimization
package,
\href{http://ab-initio.mit.edu/nlopt}{http://ab-initio.mit.edu/nlopt}.}
Like the L-M algorithm, this method requires an initial guess for the
parameter set; it also includes optional parameter limits. Unlike the
L-M algorithm, it works only with the final objective function value and
does not assume that this value must be nonnegative; thus, it is
suitable for minimizing all the fit statistics used by \imfit. It
is also as a rule less likely to be caught in local minima than the L-M
algorithm. The \textit{disadvantage} is that it is considerably
\textit{slower} than the L-M method -- roughly an order of magnitude so.

\subsubsection{Differential Evolution} 

A third alternative provided by \imfit{} is a genetic-algorithms
approach called Differential Evolution \citep[DE;][]{storn97}. This
searches the objective-function landscape using a population of
parameter-value vectors; with each ``generation'', the population is
updated by mutating and recombining some of the vectors, with new
vectors replacing older vectors if they are better-performing. DE is
designed to be -- in the context of genetic algorithms -- fast and
robust while keeping the number of adjustable \textit{algorithm}
parameters (e.g., mutation and crossover rates) to a minimum. It is the
least likely of the algorithms used by \imfit{} to become trapped in a
local minimum in the objective-function landscape: rather than starting
from a single initial guess for the parameter vector, it begins with a
set of randomly generated initial-parameter values, sampled
from the full range of allowed parameter values; in addition, the
crossover-with-mutation used to generate new parameter vectors for
successive generations helps the algorithm avoid local minima traps. Thus, in
contrast to the other algorithms, it does not require any initial
guesses for the parameter values, but \textit{does} require lower and
upper limits for all parameters. It is definitely the \textit{slowest}
of the minimization choices: about an order of magnitude slower than the
N-M simplex, and thus roughly \textit{two} orders of magnitude slower
than the L-M algorithm.

The current implementation of DE in \imfit{} uses the
``DE/rand-to-best/1/bin'' internal strategy, which controls how mutation
and crossover are done \citep{storn97}, along with a population size of
10 parameter vectors per free parameter. Since the basic DE algorithm
has no default stop conditions, \imfit{} halts the minimization when the
best-fitting value of the fit statistic has ceased to change by more
than a specified tolerance after 30 generations, or when a maximum of
600 generations is reached.

\subsubsection{Comparison and Recommendations} 

For most purposes, the default L-M method is probably the best algorithm
to use, since it is fast enough to make exploratory fitting (varying the
set of functions used, applying different parameter limits, etc.)
feasible, and also fast enough to make fitting large numbers of
individual objects in a reasonable time possible. If the problem is
relatively small (modest image size, few image functions) and the user is
concerned about possible local minima, then the N-M simplex or even the
DE algorithm can be used.

Table~\ref{tab:minimizers} provides a general comparison of the
different minimization algorithms, including the time taken for each to
find the best fit for a very simple case: a $256 \times 256$-pixel cutout of an SDSS
$r$-band image of the galaxy IC~3478, fit with a single S\'ersic function
and convolved with a $51 \times 51$-pixel PSF image. For this simple
case, the N-M approach takes $\sim 4$ times as long as the L-M method,
and the DE algorithm takes $\sim$ \textit{60} times as long. (All three
algorithms converged to the same solution, so there was no disadvantage
to using the L-M method in this case.)

\begin{deluxetable*}{llllllr}
\tablecaption{Comparison of Minimization Algorithms\label{tab:minimizers}}
\tablecolumns{7}
\tablehead{
\colhead{Algorithm} & \colhead{Initial guess} & \colhead{Bounds} & \colhead{Local-minimum} &
\colhead{Minimize $C$?} & \colhead{Speed} & \colhead{Timing} \\
                   & \colhead{required}      & \colhead{required} & \colhead{vulnerability} &
                             &                 & \colhead{example} \\
\colhead{(1)} & \colhead{(2)} & \colhead{(3)} & \colhead{(4)} & \colhead{(5)} & \colhead{(6)} & \colhead{(7)}}
\startdata
Levenberg-Marquardt (L-M)   & Yes & No  & High    & No   & Fast      & 2.2s   \\
Nelder-Mead Simplex (N-M)   & Yes & No  & Medium  & Yes  & Slow      & 9.1s  \\
Differential Evolution (DE) & No  & Yes & Low     & Yes  & Very Slow & 2m15s  \\
\enddata

\tablecomments{A comparison of the three nonlinear minimization algorithms available in
\imfit. Column 1: Algorithm name. Column 2: Notes whether an initial
guess of parameter values required. Column 3: Notes whether lower and
upper bounds on all parameter values are required. Column 4:
Vulnerability of the algorithm to becoming trapped in local minima in
the \chisquare{} (or other objective function) landscape. Column 5:
Notes whether algorithm can minimize the Cash statistic $C$ in addition
to \chisquare{} and \pmlr.
Column 6: General speed. Column 7: Approximate time taken for fitting a
$256 \times 256$ pixel SDSS galaxy image (single S\'ersic function + PSF
convolution), using a MacBook Pro with a quad-core Intel Core i7 2.3 GHz CPU (2011 model).}

\end{deluxetable*}

%
%
%
%

\subsection{Outputs and ``Goodness of Fit'' Measures} 

When \imfit{} finishes, it outputs the parameters of the best fit (along
with possible confidence intervals; see
Section~\ref{sec:confidence-intervals}) to the screen and to a text
file; it also prints the final value of the fit-statistic. The
best-fitting model image and the residual (data $-$ model) image can
optionally be saved to \textsc{fits} files as well.

For fits which minimize \chisquare, \imfit{} also prints the
\textit{reduced} \chisquare{} value, which can be used (with caution) as
an indication of the goodness of the fit. (The best-fit value of \pmlr{}
can also be converted to a reduced-\chisquare{} equivalent with the same
properties.) For fits which minimize the
Cash statistic, there is no direct equivalent to the reduced \chisquare;
the actual value of the Cash statistic does not have any directly useful
meaning by itself.

All the fit statistics (including $C$) can also be used to derive
comparative measures of how well \textit{different} models fit the same
data. To this end, \imfit{} computes two likelihood-based quantities
which can be used to compare different models. The first is the Akaike
Information Criterion (AIC, \citealt{akaike74}), which is based on an
information-theoretic approach. \Imfit{} uses the recommended,
bias-corrected version of this statistic: \begin{equation}
\mathrm{AIC_{c}} \; = \; -2 \ln \mathcal{L} \, +\,  2k \, + \,
\frac{2k(k + 1)}{n - k - 1}, \end{equation} where $\mathcal{L}$ is the
likelihood value, $k$ is the number of (free) parameters in the model
and $n$ is the number of data points. The second quantity is the
Bayesian Information Criterion (BIC, \citealt{schwarz78}), which is
\begin{equation} \mathrm{BIC} \; = \; -2 \ln \mathcal{L} \, + \, k \ln
n. \end{equation} When two or more models fit to the same data are
compared, the model with the \textit{lowest} AIC (or BIC) is preferred,
though a difference $\Delta$AIC or $\Delta$BIC of at least $\sim 6$ is
usually required before one model can be deemed clearly superior (or
inferior); see, e.g., \citet{takeuchi00} and \citet{liddle07} for
discussions of AIC and BIC in astronomical contexts, and
\citet{burnham-anderson02} for more general background. Needless to say,
all models being compared in this manner should be fit by
minimizing the same fit statistic.

\section{Confidence Intervals for Fitted Parameters}\label{sec:confidence-intervals} 

In addition to its speed, the Levenberg-Marquardt minimization algorithm
has the convenient advantage that it can automatically produce a set of
approximate, 1-$\sigma$ confidence intervals for the fitted parameters
as a side product of the minimization process; this comes from inverting
the Hessian matrix computed during the minimization process \citep[see,
e.g., Section 15.5 of][]{press92-nr}.

The other minimization algorithms available in \imfit{} do not compute
confidence intervals. Although one can, as a workaround, re-run \imfit{}
using the L-M algorithm on a solution that was found using one of the
other algorithms, this will not work if $C$ (rather than
\chisquare{} or \pmlr) was being minimized (see
Section~\ref{sec:minimization}).


An alternate method of estimating confidence intervals is provided by
bootstrap resampling \citep{efron79}. Each iteration of the resampling
process generates a new data image by sampling pixel values, with
replacement, from the original data image. (What is actually generated
inside \imfit{} is a resampled vector of pixel \textit{indices} into the
image, excluding those indices corresponding to masked pixels; the
corresponding $x, y$ coordinates and intensities then form the resampled
data.) The fit is then re-run with the best-fit parameters from the
original fit as starting values, using the L-M algorithm for
\chisquare{} and \pmlr{} minimization cases and the N-M simplex
algorithm when $C$ minimization is being done.  After $n$ iterations,
the combined set of bootstrapped parameter values is used as the
distribution of parameter values, from which properly asymmetric 68\%
confidence intervals are directly determined, along with the standard
deviation. (The 68\% confidence interval corresponds to $\pm 1$-$\sigma$
if the distribution is close to Gaussian.) 

\begin{figure}
\begin{center}
\includegraphics[scale=0.4]{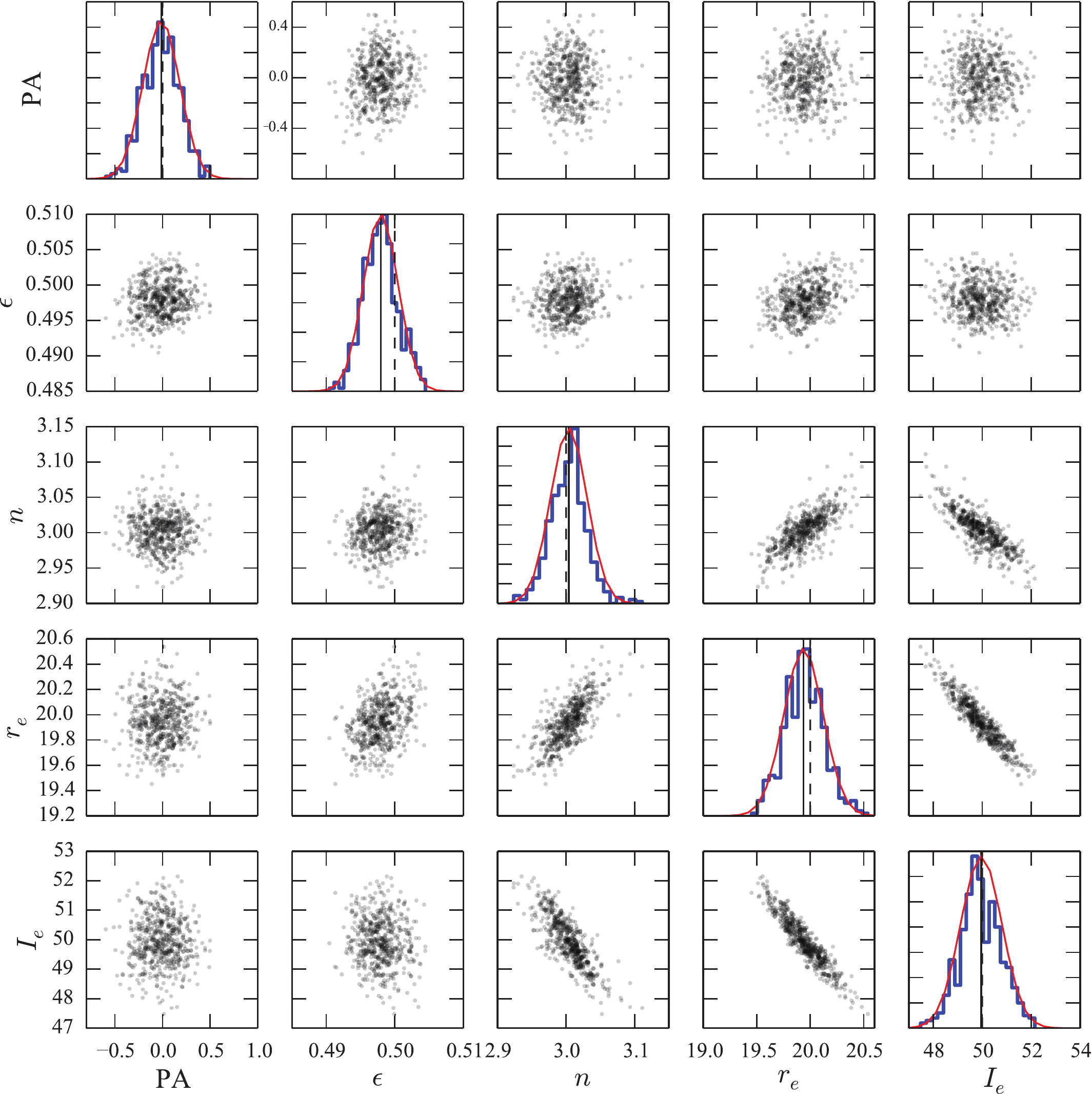}
\end{center}

\caption{Scatter-plot matrix showing bootstrap resampling analysis
of a S\'ersic fit to a simple model image with noise (500 rounds of
bootstrap resampling, using \pmlr{} for minimization). Each panel plots
the best-fit S\'eric parameter values from individual bootstrap
iterations as gray points (panels in the upper right of the plot are
rotated duplicates of those in the lower left), except for the panels
along the diagonal, which show histograms of individual parameter values
(thick blue lines). Plotted on top of the latter are Gaussians with
estimated dispersions $\sigma$ from the Levenberg-Marquardt output of the
original fit (thin
red curves). Vertical solid gray lines show the parameter values from
the fit to the original image; vertical dashed gray lines show the true
parameters of the original model.\label{fig:bootstrap}}

\end{figure}

In addition, the full set of best-fit parameters from all the
bootstrap iterations can optionally be saved to a file, which
potentially allows for more sophisticated analyses.
Figure~\ref{fig:bootstrap} shows a scatter-plot matrix comparing
parameter values for five parameters of a simple S\'ersic fit to an
image of a model S\'ersic galaxy with noise (see
Section~\ref{sec:biases} for details of the model images). One can see
that the distributions are approximately Gaussian, have dispersions
similar to those from the L-M estimates (plotted as Gaussians using red
curves), and also that certain parameter distributions are
\textit{correlated} (e.g., $n$ and $r_{e}$ or, more weakly, ellipticity
$\epsilon$ and $r_{e}$). Of course, this simple case ignores the
complexities and sources of error involved in fitting real images of
galaxies; see the Appendix for sample comparisons of L-M and bootstrap
error estimates for a small set of real-galaxy images.

The only drawback of the bootstrap-resampling approach is the cost in
time. Since bootstrap resampling should ideally use a minimum of several
hundred to one thousand or more iterations, one ends up, in effect,
re-running the fit that many times. (Some time is saved by starting each
fit with the original best-fit parameter values, since those will almost
always be close to the best-fit solution for the resampled data.)


%
%

\section{Image Functions}\label{sec:image-funcs} 

Image functions are implemented in \imfit{} as subclasses of an abstract
base class called FunctionObject. The rest of the program does not need
to know the details of the individual functions, only that they adhere
to the FunctionObject interface. This makes it relatively simple to add
new image functions to the program: write a header file and an
implementation file for the new function, add a reference to it in
another source file, and recompile the program. Further notes on how to
do this are included in the documentation.

This section describes the various default image functions that come with
\imfit. Specifications for the actual parameters (e.g., the order that
\imfit{} expects to find them in) are included in the documentation, and
a summary of all available function names and their corresponding
parameter lists can be printed using the \texttt{{-}{-}list-parameters}
command-line flag.

\subsection{2D Components}\label{sec:2d} 

Most image functions, unless otherwise noted, have two ``geometric''
parameters: the position angle PA in degrees counter-clockwise from the
vertical axis of the image\footnote{Reproducing the usual convention for
images with standard astronomical orientation, where north is up and
east is to the left.} and the ellipticity $\epsilon \; = \; 1 - b/a$,
where $a$ and $b$ are the semi-major and semi-minor axes, respectively.

In most cases, the image function internally converts the ellipticity to
an axis ratio $q = b/a$ ($= 1 - \epsilon$) and the position angle to an angle
relative to the image $x$-axis $\theta$ ($= {\rm PA} + 90\arcdeg$), in
radians. Then for each input pixel (or subpixel if pixel subsampling is
being done) with image coordinates $(x, y)$ a scaled radius is computed
as
\begin{equation}
r \; = \; \left(  x_{p}^{2} \: + \: \frac{y_{p}^{2}}{q^{2}} \right)^{1/2},
\end{equation}
where $x_{p}$ and $y_{p}$ are coordinates in the reference frame centered on the
image-function center $(x_{0},y_{0})$ and rotated to its position angle:
\begin{eqnarray}\label{eqn:xy}
x_{p} \; & = & \; (x - x_{0}) \, \cos\theta \, + \, (y - y_{0}) \, \sin\theta \\
y_{p} \; & = & \; -(x - x_{0}) \, \sin\theta \, + \, (y - y_{0}) \, \cos\theta \nonumber
\end{eqnarray}
This scaled radius is then used to compute the actual intensity, using
the appropriate 1-D intensity function (see descriptions of individual
image functions, below).

Pure circular versions of any of these functions can be had by
specifying that the ellipticity parameter is fixed, with a value of 0.
Some functions (e.g., EdgeOnDisk) have only the position angle as a
geometric parameter, and instead of computing a scaled radius, convert
the pixel coordinates to corresponding $r$ and $z$ values in the rotated
2D coordinate system of the model function.

%
%

\subsubsection{FlatSky}

This is a very basic function which produces a uniform background:
$I(x,y) = I_{\rm sky}$ for all pixels. Unlike most image functions, it
has \textit{no} geometric parameters.

\subsubsection{Gaussian}

This is an elliptical 2D Gaussian function, with central surface brightness $I_{0}$
and dispersion $\sigma$. The intensity
profile is given by
\begin{equation}
I(r) \, = \, I_{0} \, \exp \left(-\frac{r^2}{2 \sigma^2} \right).
\end{equation}

\subsubsection{Moffat}

This is an elliptical 2D function with a \citet{moffat69} function for the
surface brightness profile, with parameters for the central surface
brightness $I_{0}$, full-width half-maximum (FWHM), and the shape parameter $\beta$.
The intensity profile is given by
\begin{equation}
I(r) \; = \; \frac{I_{0}  }{(1 \, + \, (r/\alpha)^{2})^{\beta} },
\end{equation}
where $\alpha$ is defined as
\begin{equation}
\alpha \; = \; \frac{ {\rm FWHM}}{2 \sqrt{2^{1/\beta} - 1}}.
\end{equation}
In practice, FWHM describes the overall width of the profile, while $\beta$ describes the
strength of the wings: lower values of $\beta$ mean more intensity in the wings
than is the case for a Gaussian (as $\beta \rightarrow \infty$, the Moffat profile
converges to a Gaussian).

The Moffat function is often a good approximation to typical telescope
PSFs (see, e.g., \citealt{trujillo-moffat01}), and \makeimage{} can
easily be used to generate Moffat PSF images.

\subsubsection{Exponential and Exponential\_GenEllipse}

The Exponential function is an elliptical 2D exponential function, with
parameters for the central surface brightness $I_{0}$ and the
exponential scale length $h$. The intensity profile is given by
\begin{equation}
I(r) \; = \; I_{0} \, \exp(-r/h);
\end{equation}
together with the position angle and ellipticity, there are a total of
four parameters. This is a good default for galaxy disks seen at
inclinations $\la 80\arcdeg$, though the majority of disk galaxies have
profiles which are more complicated than a simple exponential
\citep[e.g.,][]{gutierrez11}.

The Exponential\_GenEllipse function is identical to the
Exponential function except for allowing the use of generalized ellipses
(with shapes ranging from ``disky'' to pure elliptical to ``boxy'') for the
isophote shapes. Following \citet{athanassoula90} and \citet{peng02},
the shape of the elliptical isophotes is controlled by the $c_{0}$
parameter, such that a generalized ellipse with ellipticity $= 1 - b/a$
is described by
\begin{equation}
\left( \frac{|x|}{a} \right)^{c_{0} \,+\, 2} \!\! + \; \left( \frac{|y|}{b} \right)^{c_{0} \,+\, 2}  = \; 1,
\end{equation}
where $|x|$ and $|y|$ are distances from the ellipse center in the
coordinate system aligned with the ellipse major axis ($c_{0}$
corresponds to $c - 2$ in the original formulation of
\nocite{athanassoula90}Athanassoula et al.). Thus, values of $c_{0} < 0$
correspond to disky isophotes, while values $> 0$ describe boxy
isophotes; $c_{0} = 0$ for a perfect ellipse.

%



\subsubsection{Sersic and Sersic\_GenEllipse}

This pair of related functions is analogous to the Exponential and
Exponential\_GenEllipse pair above, except that the intensity profile is
given by the \citet{sersic68} function:
\begin{equation}
I(r) \; = \; I_{e} \: \exp \left\{ -b_{n} \left[ \left( \frac{r}{r_{e}} \right)^{1/n} \! - \: 1 \right] \right\},
\end{equation}
where $I_{e}$ is the surface brightness at the effective (half-light) radius $r_{e}$
and $n$ is the index controlling the shape of the intensity profile. The
value of $b_{n}$ is formally given by the solution to the transcendental equation
\begin{equation}
\Gamma(2 n) \; = \; 2 \gamma(2n, b_{n}),
\end{equation}
where $\Gamma(a)$ is the gamma function and $\gamma(a, x)$ is the
incomplete gamma function. However, in the current implementation
$b_{n}$ is calculated via the polynomial approximation of
\citet{ciotti99} when $n > 0.36$ and the approximation of
\citet{macarthur03} when $n \leq 0.36$.

The S\'ersic profile is equivalent to the de Vaucouleurs ($r^{1/4}$)
profile when $n = 4$, to an exponential when $n = 1$, and to a Gaussian
when $n = 0.5$; it has become the de facto standard for fitting the
surface-brightness profiles of elliptical galaxies and bulges.
Though the empirical justification for doing so is rather limited,
the combination a S\'ersic profile with $n < 1$ and isophotes with a boxy shape
is often used to represent bars when fitting images of disk galaxies. In
addition, the combination of boxy isophotes and high $n$ values may be
appropriate for modeling luminous boxy elliptical galaxies.


%


\subsubsection{Core-Sersic}

This function generates an elliptical 2D function where the major-axis
intensity profile is given by the Core-S{\'e}rsic model
\citep{graham03,trujillo04}, which was designed to fit the profiles of
so-called ``core''  galaxies
\citep[e.g.,][]{ferrarese06b,richings11,dullo12,dullo13,rusli13b}. It consists
of a S\'ersic profile (parameterized by $n$ and $r_{e}$) for radii $>$
the break radius $r_{b}$ and a single power law with index $-\gamma$ for
radii $< r_{b}$. The transition between the two regimes is mediated by
the dimensionless parameter $\alpha$: for low values of $\alpha$, the
transition is very gradual and smooth, while for high values of $\alpha$
the transition becomes very abrupt (a perfectly sharp transition can be
approximated by setting $\alpha$ equal to some large number, such as
100).  The intensity profile is given by
\begin{equation}
I(r) \; = \; I_{b}
\left[1 + \left( \frac{r_b}{r} \right)^{\alpha} \right]^{\gamma/\alpha}
\exp \left[ -b \left( \frac{r^\alpha+r_b^\alpha}{r_e^\alpha} \right)^{1/(n\alpha)}
\right],
\end{equation}
where $b$ is the same as $b_{n}$ for the S{\'e}rsic function.

The overall intensity scaling is set by $I_{b}$,
the intensity at the break radius $r_{b}$:
\begin{equation}
I_{b} \; = \; I_b \; 2^{-\gamma/\alpha} \exp[\, b \, 2^{1/\alpha n} \,
(r_b/r_e)^{1/n}].
\end{equation}



\subsubsection{BrokenExponential}\label{sec:brokenexp}

\begin{figure}
\begin{center}
\includegraphics[scale=0.47]{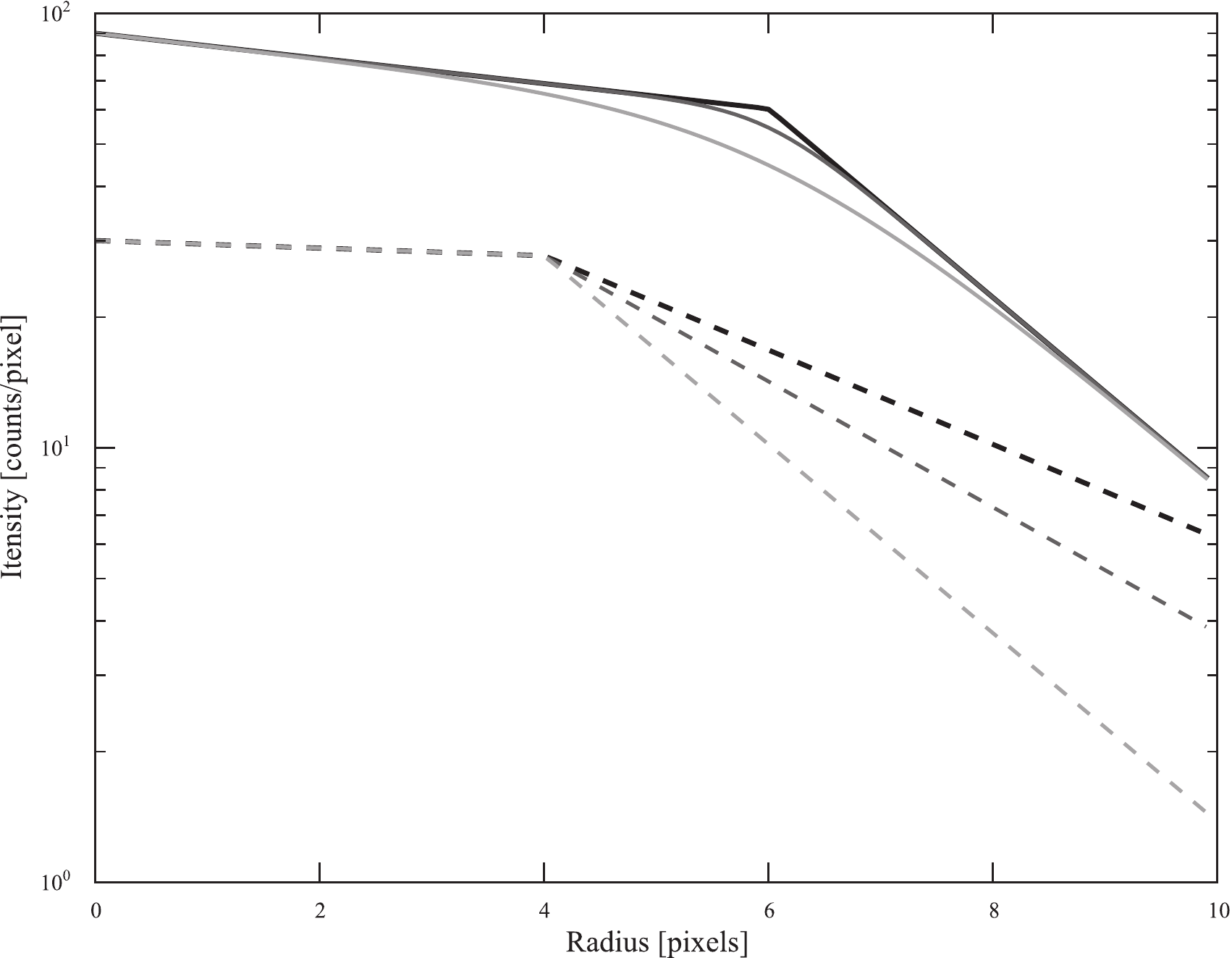}
\end{center}

\caption{Examples of the ``broken-exponential'' surface-brightness
profile used by the BrokenExp and BrokenExp3D image functions. The upper
(solid) curves show a profile with inner and outer scale lengths $h_{1}
= 15$ and $h_{2} = 2$ pixels, respectively, break radius = 6 pixels, and
varying values of $\alpha$ (black = 100, medium gray = 3, light gray =
1). The lower (dashed) curves show the effects of varying the outer
scale length only ($h_{2} = 4$, 3, 2 pixels).\label{fig:brokenexp}}

\end{figure}

This is similar to the Exponential function, but it has \textit{two}
exponential radial zones (with different scalelengths) joined by a transition region
at $R_{b}$ of variable sharpness:
\begin{equation}
	I(r) \; = \; S \, I_{0} \, e^{-\frac{r}{h_{1}}} [1 + e^{\alpha(r \, - \,
	R_{b})}]^{\frac{1}{\alpha} (\frac{1}{h_{1}} \, - \, \frac{1}{h_{2}})},
\end{equation}
where $I_{0}$ is the central intensity of the inner exponential, $h_{1}$
and $h_{2}$ are the inner and outer exponential scale lengths, $R_{b}$
is the break radius, and $\alpha$ parameterizes the sharpness of the
break.  Low values of $\alpha$ mean very smooth, gradual breaks, while
high values correspond to abrupt transitions.  $S$ is a scaling factor,\footnote{As pointed 
out by \citet{munoz-mateos13},
the original definition of this factor in Eqn.~6 of \citet{erwin08} contained a typo.}
given by
\begin{equation}
  S \; = \; (1 + e^{-\alpha R_{b}})^{- \frac{1}{\alpha} (\frac{1}{h_{1}} \, - 
  \, \frac{1}{h_{2}})};
\end{equation}
see Figure~\ref{fig:brokenexp} for examples. Note that the parameter $\alpha$ 
has units of length$^{-1}$ (pixels$^{-1}$ for the specific case of \imfit).

The 1D form of this profile \citep{erwin08} was designed to fit the
surface-brightness profiles of disks which are not single-exponential:
e.g., disks with truncations or antitruncations \citep{erwin05,erwin08,munoz-mateos13}.

\subsubsection{GaussianRing}\label{sec:gaussian-ring}

This function creates an elliptical ring with a Gaussian radial profile, centered
at $r = R_{\rm ring}$ along the major axis.
\begin{equation}
I(r) \, = \, I_{0} \, \exp \left(-\frac{(r \, - \, R_{\rm ring})^2}{2 \sigma^2} \right).
\end{equation}

See Figure~\ref{fig:rings} for an example.



\subsubsection{GaussianRing2Side}

\begin{figure}
\begin{center}
\includegraphics[scale=0.6]{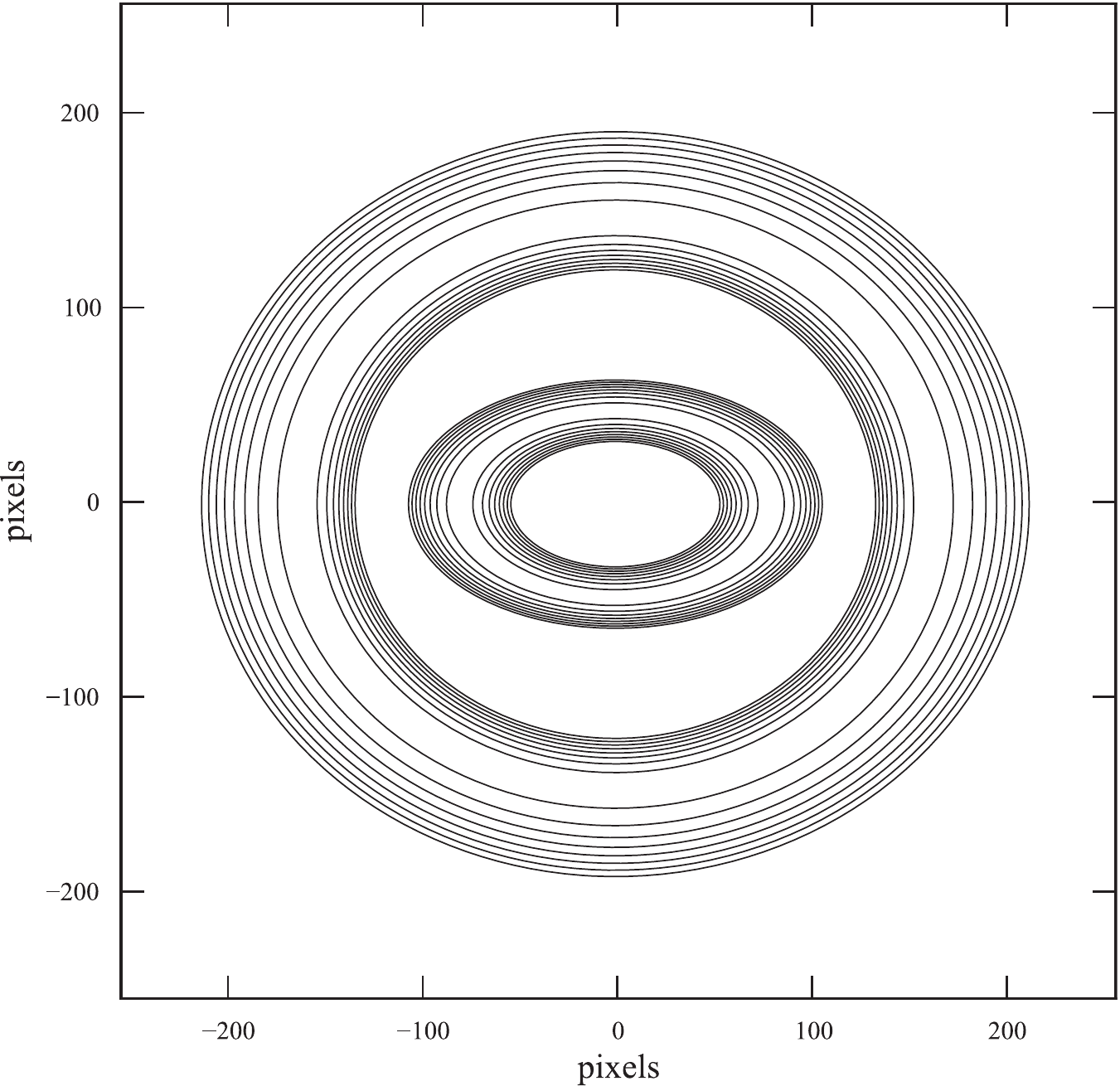}
\end{center}

\caption{Logarithmically scaled isophotes for examples of the Gaussian
ring image functions. The inner, more elliptical ring was generated by
the GaussianRing function, with an ellipticity of 0.4, semi-major axis
of 80 pixels, and $\sigma = 10$ pixels. The larger, rounder ring is an
example of the GaussianRing2Side function, with an ellipticity of 0.1, a
semi-major axis of 160 pixels, $\sigma_{\rm in} = 10$ pixels, and
$\sigma_{\rm out} = 20$ pixels. \label{fig:rings}}

\end{figure}

This function is similar to GaussianRing, except that it uses an
asymmetric Gaussian, with different values of $\sigma$ for $r < R_{\rm
ring}$ and $r > R_{\rm ring}$). That is, the profile behaves as
\begin{equation}
I(r) \, = \, I_{0} \, \exp \left(-\frac{(r \, - \, R_{\rm ring})^2}{2 \sigma_{\rm in}^2} \right)
\end{equation}
for $r < R_{\rm ring}$, and 
\begin{equation}
I(r) \, = \, I_{0} \, \exp \left(-\frac{(r \, - \, R_{\rm ring})^2}{2 \sigma_{\rm out}^2} \right)
\end{equation}
for $a > R_{\rm ring}$; see Figure~\ref{fig:rings} for an example.



\subsubsection{EdgeOnDisk}\label{sec:edgeondisk}

This function provides the analytic form for a perfectly edge-on disk
with a radial exponential profile, using the Bessel-function solution of
\citet{vanderkruit81a} for the radial profile. Although it is common to
assume that the vertical profile for galactic disks follows a ${\rm
sech}^2$ function, based on the self-gravitating isothermal sheet model
of \citet{spitzer42}, \citet{vanderkruit88} suggested a more generalized
form for this, one which enables the profile to range from ${\rm
sech}^2$ at one extreme to exponential at the other:
\begin{equation}
L(z) \; \propto \; {\rm sech}^{2/n}(n z / (2 z_{0})),
\end{equation}
with $z$ the vertical coordinate and $z_{0}$ the vertical scale height.
The parameter $n$ produces a ${\rm sech}^2$ profile when $n = 1$, ${\rm
sech}$ when $n = 2$, and converges to an exponential as $n \rightarrow
\infty$. See \citet{degrijs97} for examples of fitting the vertical
profiles of edge-on galaxy disks using this formula, and
\citet{yoachim06} for examples of 2D fitting of edge-on galaxy images.

In a coordinate system aligned with the edge-on disk, $r$ is the
distance from the minor axis (parallel to the major axis) and $z$ is the
perpendicular direction, with $z = 0$ on the major axis. (The latter
corresponds to height $z$ from the galaxy midplane.) The intensity at
$(r,z)$ is given by
\begin{equation}
I(r,z) \; = \; \mu(0,0) \; (r/h) \; K_{1}(r/h) \;\, {\rm sech}^{2/n} (n \, z/(2 \, z_{0}))
\end{equation}
where $h$ is the exponential scale length in the disk plane, $z_{0}$ is
the vertical scale height, and $K_{1}$ is the modified Bessel function
of the second kind.
The central surface brightness $\mu(0,0)$ is given by
\begin{equation}
\mu(0,0) \; = \;  2 \, h \, L_{0},
\end{equation}
where $L_{0}$ is the central luminosity \textit{density} (see
\citealt{vanderkruit81a}). Note that $L_{0}$ is the actual input parameter
required by the function; $\mu(0,0)$ is calculated internally.

\begin{figure*}
\begin{center}
\includegraphics[scale=0.95]{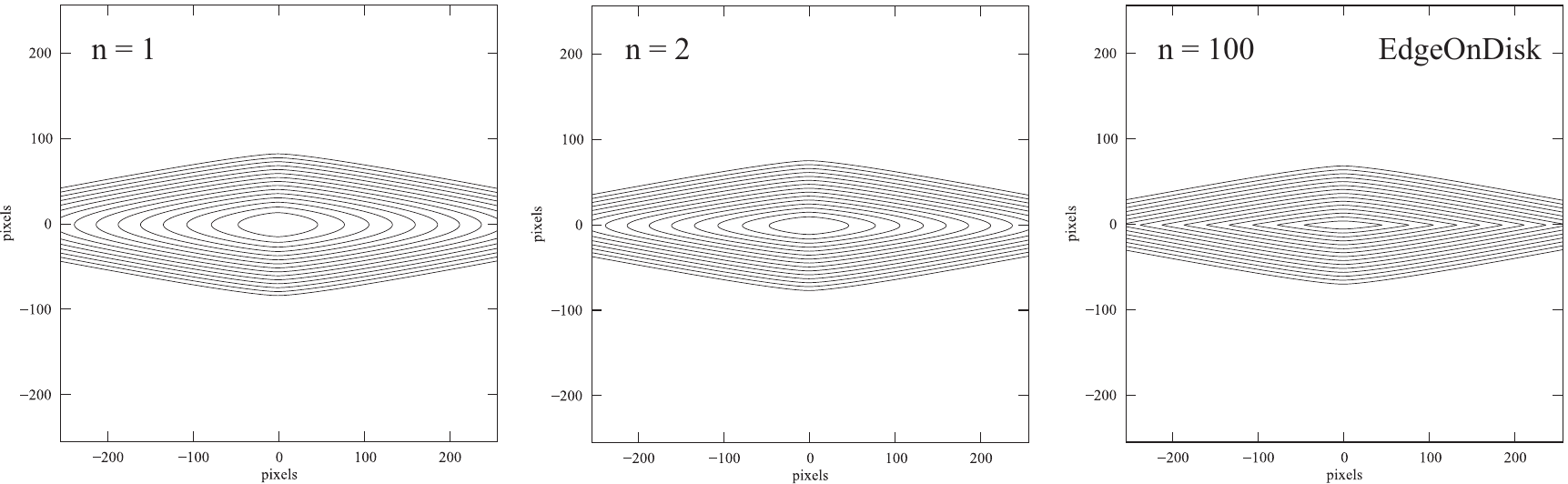}
\end{center}

\caption{Examples of the EdgeOnDisk image function, which uses the
analytic Bessel-function solution of \citet{vanderkruit81a} for a
perfectly edge-on exponential disk, combined with the generalized ${\rm sech}^{2/n}$
vertical profile of \citet{vanderkruit88}. All panels show models with radial
and vertical scale lengths $h = 50$ and $z_{0} = 10$ pixels,
respectively. From left to right, the panels show images with vertical
${\rm sech}^{2/n}$ profiles having $n = 1$ (${\rm sech}^2$ profile), 2
(${\rm sech}$ profile), and 100 ($\approx$ exponential
profile).\label{fig:edgeondisk}}

\end{figure*}

The result is a function with five parameters: $L_{0}$, $h$, $z_0$, $n$, and
the position angle; Figure~\ref{fig:edgeondisk} shows three examples with differing
vertical profiles parameterized by $n = 1$, 2, and 100.

\subsubsection{EdgeOnRing}

\begin{figure*}
\begin{center}
\includegraphics[scale=0.95]{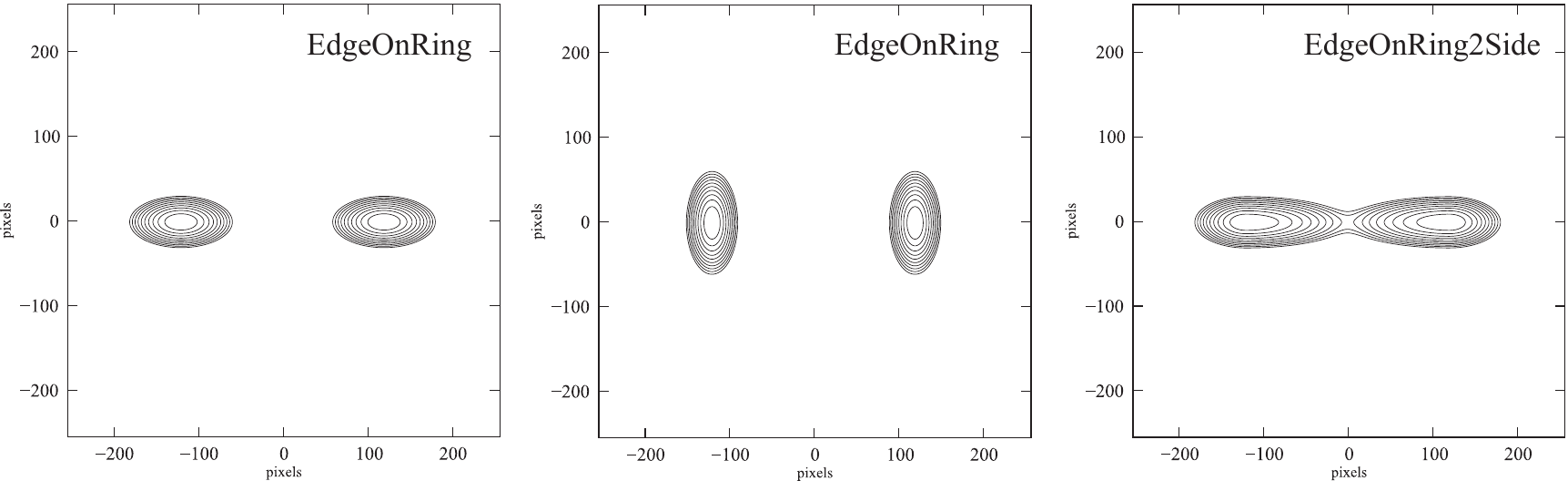}
\end{center}

\caption{Examples of the EdgeOnRing (left and middle panels) and
EdgeOnRing2Side (right panel) image functions, which provide simple approximations
for rings seen edge-on. All rings have
a radius of 120 pixels. The left-hand panel shows a ring with radial and
vertical Gaussian widths of 20 and 10 pixels, respectively; the middle
panel shows a model with the radial and vertical widths exchanged. The
right-hand panel shows an example of the EdgeOnRing2Side function, where
the radial scales are $\sigma = 40$ pixels on the inside and 20 pixels
on the outside; the vertical scale is 10 pixels. \label{fig:edgeonring}}

\end{figure*}

This is a  simplistic model for an edge-on ring, using two offset
subcomponents located at distance $R_{\rm ring}$ from the center of the
function block. Each subcomponent (i.e., each side of the ring) is a 2D
Gaussian with central surface brightness $I_{0}$ and dispersions of
$\sigma_{r}$ in the radial direction and $\sigma_{z}$ in the vertical
direction. It has five parameters: $I_{0}$, $R_{\rm ring}$,
$\sigma_{r}$, $\sigma_{z}$, and the position angle. See
Figure~\ref{fig:edgeonring} for examples of this function.

A potentially more correct (though computationally more expensive) model for
a ring seen edge-on ring -- or at other inclinations -- is provided
by the GaussianRing3D function, below.

\subsubsection{EdgeOnRing2Side}

This is a slightly more sophisticated variant of EdgeOnRing, where the
radial profile for the two components is an asymmetric Gaussian, as in
the case of the GaussianRing2Side function, above: the inner ($|r| <
R_{\rm ring}$) side of each component is a Gaussian with radial
dispersion $\sigma_{r,{\rm in}}$, while the outer side has radial
dispersion $\sigma_{r,{\rm out}}$. It thus has six parameters: $I_{0}$,
$R_{\rm ring}$, $\sigma_{r,{\rm in}}$, $\sigma_{r,{\rm out}}$,
$\sigma_{z}$, and the position angle. See the right-hand panel of
Figure~\ref{fig:edgeonring} for an example.

\subsection{3D Components}\label{sec:image-funcs-3d} 

All image functions in \imfit{} produce 2D surface-brightness output.
However, there is nothing to prevent one from creating a function which
does something quite complicated in order to produce this output. As an
example, \imfit{} includes three image functions which perform
line-of-sight integration through 3D luminosity-density models, in order
to produce a 2D projection.

These functions assume a symmetry plane (e.g., the disk plane for a disk
galaxy) which is inclined with respect to the line of sight; the
inclination is defined as the angle between the line of sight and the
normal to the symmetry plane, so that a face-on system has $i =
0\arcdeg$ and an edge-on system has $i = 90\arcdeg$. For inclinations $>
0\arcdeg$, the orientation of the line of nodes (the intersection
between the symmetry plane and the sky plane) is specified by a
position-angle parameter $\theta$. Instead of a 2D surface-brightness
specification (or 1D radial surface-brightness profile), these functions
specify a 3D luminosity density $j$, which is numerically integrated along the line of
sight $s$ for each pixel of the model image:
\begin{equation}
I(x,y) \; = \; \int_{-S}^{S} j(s) \:\mathrm{d}s.
\end{equation}

%


To carry out the integration for a pixel located at $(x,y)$ in the image
plane, the coordinates are first transformed to a rotated image plane
system $(x_{p},y_{p})$ centered on the coordinates of the component
center $(x_{0},y_{0})$, where the line of nodes lies along the $x_{p}$
axis (cf.\ Eqn.~\ref{eqn:xy} in Section~\ref{sec:2d}):
\begin{eqnarray*}
x_{p} \; & = & \; (x - x_{0}) \, \cos\theta \, + \, (y - y_{0}) \, \sin\theta \\
y_{p} \; & = & \; -(x - x_{0}) \, \sin\theta \, + \, (y - y_{0}) \, \cos\theta 
\end{eqnarray*}
with $\theta$ being the angle between the line of nodes and the image
$+x$ axis (as in the case of the 2D functions, the actual user-specified
parameter is $\mathrm{PA} = \theta - 90$, which is the angle between the
line of nodes and the $+y$ axis).

\begin{figure}
\begin{center}
\includegraphics[scale=0.7]{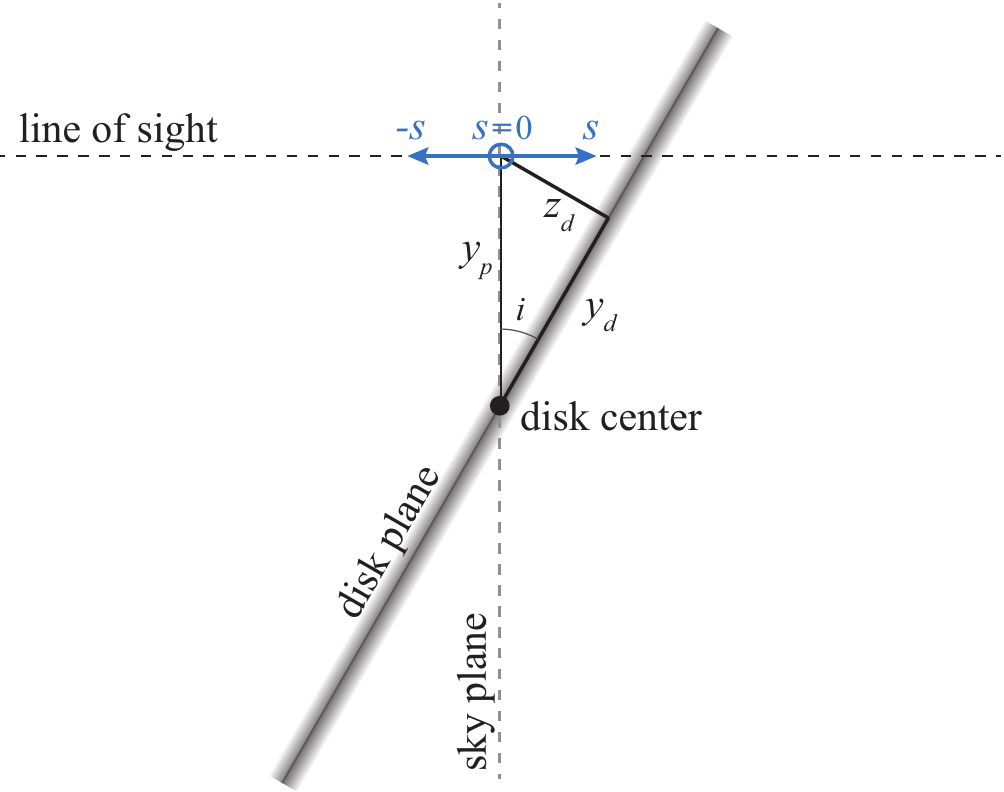}
\end{center}

\caption{A simplified illustration of how line-of-sight integration is
handled for 3D image functions. Here, an axisymmetric ExponentialDisk3D
component is inclined at angle $i$ with respect to the line of sight,
with the line of nodes rotated to lie along the sky-plane $x_{p}$ axis,
perpendicular to the page; the disk center is by construction at the
intersection of the disk plane and the sky plane. For a pixel with
sky-plane coordinates $(x_{p},y_{p})$, the luminosity-density is
integrated along the line of sight (variable $s$, with $s = 0$ at the
sky plane). For each value of $s$ used by the integration routine, the
luminosity density is computed based on the corresponding values of
radius $r = (x_{d}^{2} + y_{d}^{2})^{1/2}$ and height $z_{d}$ in the disk's native coordinate system.
\label{fig:3d-integration}}

\end{figure}

The line-of-sight coordinate $s$ is then defined so that $s = 0$ in the
sky plane (an instance of the image plane located in 3D space so that it
passes through the center of the component), corresponding to 
\begin{eqnarray*}
x_{d,0} & \; = & \; x_{p} \\
y_{d,0} & \; = & \; y_{p} \cos i \\
z_{d,0} & \; = & \; y_{p} \sin i
\end{eqnarray*}
in the component's native $(x_{d},y_{d},z_{d})$ Cartesian coordinate system. A location 
at $s$ along the line of sight then maps into the component coordinate system as
\begin{eqnarray}\label{eqn:3d}
y_{d} & \; = & \; y_{d,0} \, + \, s \, \sin i \nonumber \\
z_{d} & \; = & \; z_{d,0} \, - \, s \: \cos i,
\end{eqnarray}
with $x_{d} = x_{d,0} = x_{p}$ by construction. The luminosity-density value is
then $j(s) = j(x_{d},x_{d},z_{d})$. See Figure~\ref{fig:3d-integration} for
a side-on view of this arrangement.

Although a fully correct integration would run from $s = -\infty$
to $\infty$, in practice the limit $S$ is some large multiple of the
component's normal largest scale size (e.g., 20 times the horizontal
disk scale length $h$), to limit the possibility of numerical integration
mishaps.

\subsubsection{ExponentialDisk3D}\label{sec:expdisk3d}

This function implements a 3D luminosity density model for an
axisymmetric disk where the radial profile of the luminosity density is
an exponential  and the vertical profile follows the ${\rm sech}^{2/n}$
function of \citet{vanderkruit88} (see the discussion of the EdgeOnDisk
function in Section~\ref{sec:edgeondisk}).  The line-of-sight
integration is done numerically, using functions from the GNU Scientific
Library.

In a cylindrical coordinate system $(r, z)$ aligned with the disk (where the disk
midplane has $z = 0$), the luminosity density $j(r,z)$ at radius $r$ from 
the central axis and
at height $z$ from the midplane is given by
\begin{equation}
j(r,z) \; = \; J_{0} \; \exp(-r/h) \; {\rm sech}^{2/n} (n \, z/(2 \, z_{0}))
\end{equation}
where $h$ is the exponential scale length in the disk plane, $z_{0}$ is the vertical
scale height, $n$ controls the shape of the vertical distribution, and $J_{0}$ is 
the central luminosity density. Note that in the context of the introductory
discussion above, $z = z_{d}$ and $r = (x_{d}^{2} + y_{d}^{2})^{1/2}$.

\begin{figure*}
\begin{center}
\includegraphics[scale=0.95]{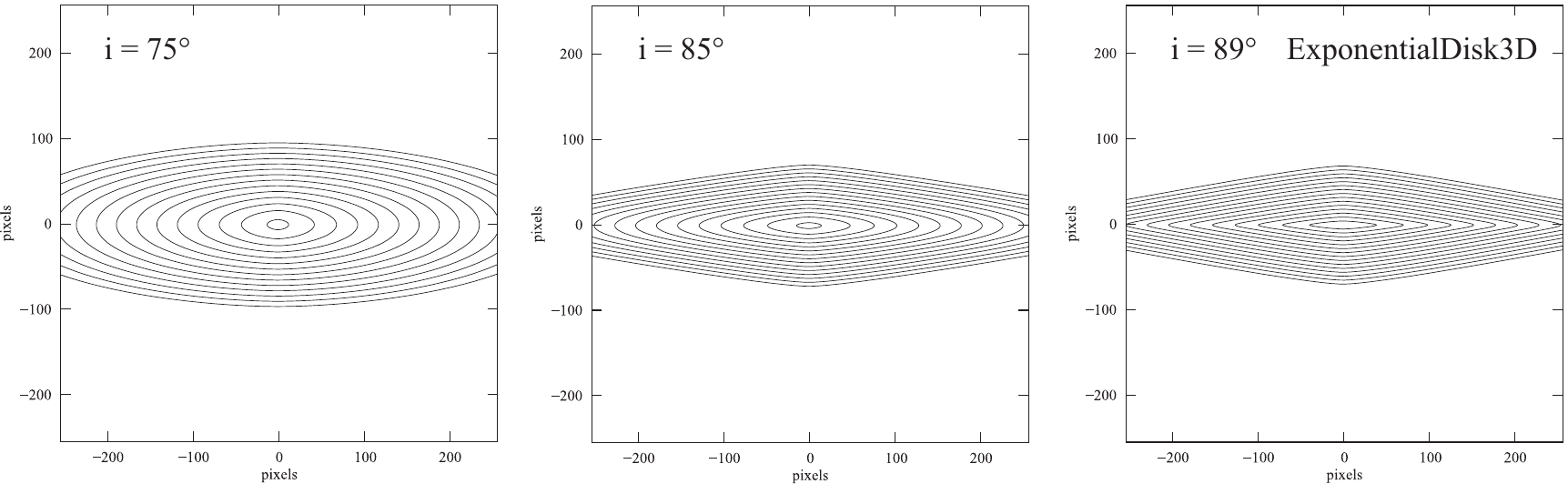}
\end{center}

\caption{Examples of the ExponentialDisk3D image function, which uses
line-of-sight integration through a 3D luminosity-density model of a
disk with radial exponential profile and vertical ${\rm sech}^{2/n}$
profile. All panels show the same model, with radial and vertical scale
lengths $h = 50$ and $z_{0} = 10$ pixels, respectively, and a vertical
exponential profile ($n = 100$). From left to right, the panels show
projections with inclinations of 75\arcdeg, 85\arcdeg, and
89\arcdeg; compare the last panel with the right-hand panel in Figure~\ref{fig:edgeondisk}.\label{fig:expdisk3d}}

\end{figure*}

Figure~\ref{fig:expdisk3d} shows three views of the same model,
at inclinations of 75\arcdeg, 85\arcdeg, and 89\arcdeg; the latter is almost identical
to the image produced by the analytic EdgeOnDisk with the same radial and vertical
parameters (right-hand panel of Figure~\ref{fig:edgeondisk}).

\subsubsection{BrokenExponentialDisk3D}

This function is identical to the ExponentialDisk3D function, except
that the radial part of the luminosity density function is given by the
broken-exponential profile used by the (2D) BrokenExponential function,
above (Section~\ref{sec:brokenexp}). Thus, the luminosity density 
$j(r,z)$ at radius $r$ from the central axis and at height $z$ from the
midplane is given by
\begin{equation}
j(r,z) \; = \; I_{\rm rad}(r) \: {\rm sech}^{2/n} (n \, z/(2 \, z_{0}))
\end{equation}
where $z_{0}$ is the vertical
scale height, and the radial part is given by
\begin{equation}
	I_{\rm rad}(r) \; = \; S \, J_{0} \, e^{-\frac{r}{h_{1}}} [1 + e^{\alpha(r \, - \,
	R_{b})}]^{\frac{1}{\alpha} (\frac{1}{h_{1}} \, - \, \frac{1}{h_{2}})},
\end{equation}
with $J_{0}$ being the central luminosity density and the rest of the
parameters as defined for BrokenExponential function (Section~\ref{sec:brokenexp}).

\subsubsection{GaussianRing3D}

This function creates the projection of a 3D elliptical ring, seen at an
arbitrary inclination.  The ring has a luminosity density
with a radial Gaussian profile (centered at $a_{\rm ring}$ along
ring's major axis, with in-plane width $\sigma$) and a vertical
exponential profile (with scale height $h_{z}$). The ring can be
imagined as residing in a plane which has its line of nodes
at angle $\theta$ and inclination $i$ (as for the
ExponentialDisk3D function, above); within this plane, the ring's major
axis is at position angle $\phi$ relative to the
perpendicular to the line of nodes. To derive the correct luminosity
densities for the line-of-sight integration, the component coordinate
values $x_{d}, y_{d}, z_{d}$ from Equation~\ref{eqn:3d} are transformed
to a system rotated about the normal to the ring plane, where the ring's major 
axis is along the $x_{\rm ring}$ axis:
\begin{eqnarray*}
x_{\rm ring} & \; = \; & x_{d} \cos(\phi) \, + \, y_{d} \sin(\phi) \\
y_{\rm ring} & \; = \; & -x_{d} \sin(\phi) \, + \, y_{d} \cos(\phi) \\
z_{\rm ring} & \; = \; & |z_{d}| .
\end{eqnarray*}

\begin{figure*}
\begin{center}
\includegraphics[scale=0.95]{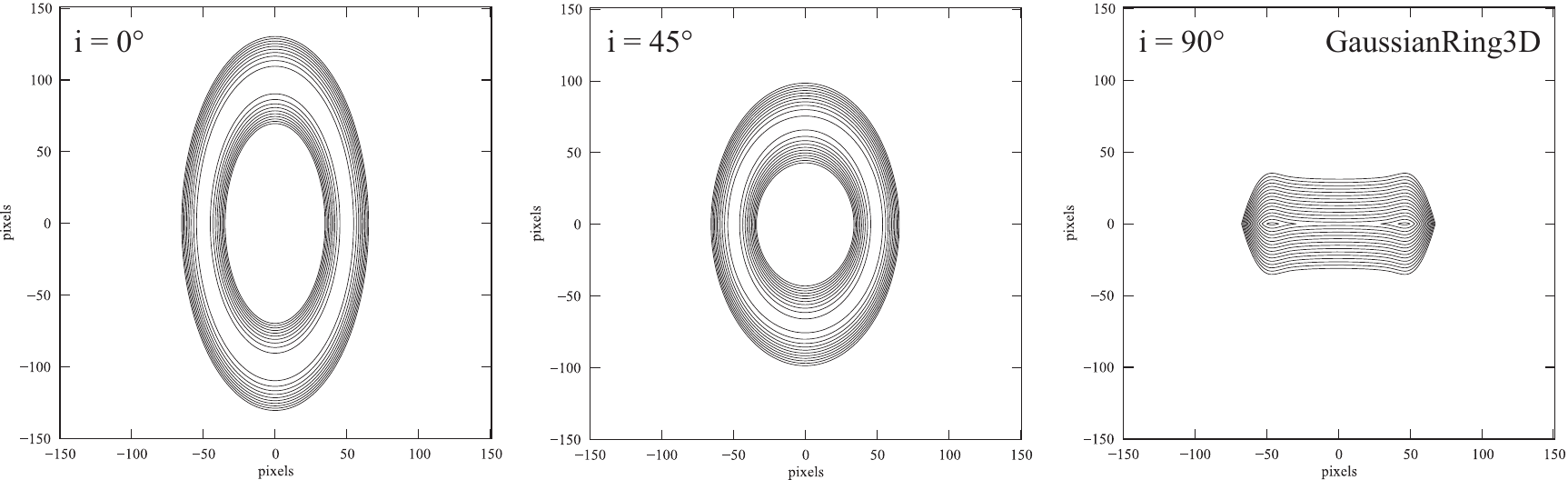}
\end{center}

\caption{Examples of the GaussianRing3D image function, which uses
line-of-sight integration through a 3D luminosity-density model of an
elliptical ring with Gaussian radial and exponential vertical profiles.
This particular ring has an intrinsic (in-plane) ellipticity $= 0.5$,
semi-major axis $= 100$ pixels, Gaussian radial width $\sigma = 10$ pixels,
and exponential scale height $h_{z} = 5$ pixels. From left to right,
panels show face-on, $i = 45\arcdeg$, and edge-on views.\label{fig:ring3d}}

\end{figure*}

Figure~\ref{fig:ring3d} shows the same
GaussianRing3D component (with ellipticity = 0.5) seen at three different
inclinations.

\section{Programming Notes}\label{sec:programming} 

\Imfit{} is written in standard C++, and should be compilable with any
modern compiler; it has been tested with GCC versions 4.2 and 4.8 on
Mac~OS~X and GCC version 4.6 on Ubuntu Linux systems. It makes use of
several open-source libraries, two which are required (CFITSIO and FFTW)
and two which are optional but recommended (NLopt and the GNU Scientific
Library). \Imfit{} also uses the Python-based
SCons\footnote{\href{http://www.scons.org}{http://www.scons.org}} build system
and CxxTest\footnote{\href{http://cxxtest.com}{http://cxxtest.com}} for unit tests.

Since the slowest part of the fitting process is almost always computing
the model image, \imfit{} is written to take advantage of OpenMP
compiler extensions; this allows the computation of the model image to be
parceled out into
multiple threads, which are then allocated among available processor cores on
machines with multiple shared-memory CPUs (including single CPUs with
multiple cores). As an example of how effective this can be, tests on a
MacBook Pro with a quad-core i7 processor, which has a total of eight
virtual threads available, show that basic computation of large images
(without PSF convolution) is sped up by a factor of $\sim 6$ when OpenMP
is used. Even when the overhead of an actual fit is included, the total
time to fit a four-component model with 21 free parameters (without
PSF convolution) to a $500 \times 500$-pixel image is reduced by a
factor of $\sim 3.8$.

Additional computational overhead is imposed when one convolves a model
image with a PSF. To mitigate this, \imfit{} uses the FFTW library to
compute the necessary Fourier transforms. This is one of the fastest FFT
libraries available, and it can be compiled with support for multiple
threads. When the same $500 \times 500$-pixel image fit mentioned above
is done including convolution with a $35 \times 35$-pixel PSF image, the
total time drops from $\sim 280$s without any multi-threading to $\sim
120$s when just the FFT computation is multi-threaded, and down to $\sim
50$s when OpenMP threading is enabled as well.

Multithreading can always be reduced or turned off using a command-line
option if one does not wish to use all available CPU cores for a given
fit.



\section{Sample Applications}\label{sec:examples} 

\Imfit{} has been used for several different astronomical applications,
including preliminary work on the EUCLID photometric pipeline
\citep{kummel13}, testing 1D convolution code used in
the analysis of core galaxies \citep{rusli13b}, fitting kinematically
decomposed components of the galaxy NGC~7217 \citep{fabricius14}, 
determining the PSF for Data Release 2 of the CALIFA survey \citep{califa-dr2},
and separation of bulge and disk components for dynamical modeling of black
hole masses in nearby S0 and spiral galaxies (Erwin et al., in prep). 

In this section I present two relatively simple examples of using
\imfit{} to model images of real galaxies. The first case considers a
moderately inclined spiral galaxy with a prominent ring surrounding a
bar, where use of a separate ring component considerably improves the
fit. The second case is an edge-on disk galaxy with both thin and thick
disks; I show how this can be fit using both the analytic pure-edge-on
disk component (EdgeOnDisk; Section~\ref{sec:edgeondisk}) and the 3D
luminosity-density model of an exponential disk (ExponentialDisk3D;
Section~\ref{sec:expdisk3d}).

\subsection{PGC 35772: Disk, Bar, and Ring}\label{sec:haggis-fit} 

\begin{figure*}
\begin{center}
\includegraphics[scale=0.8]{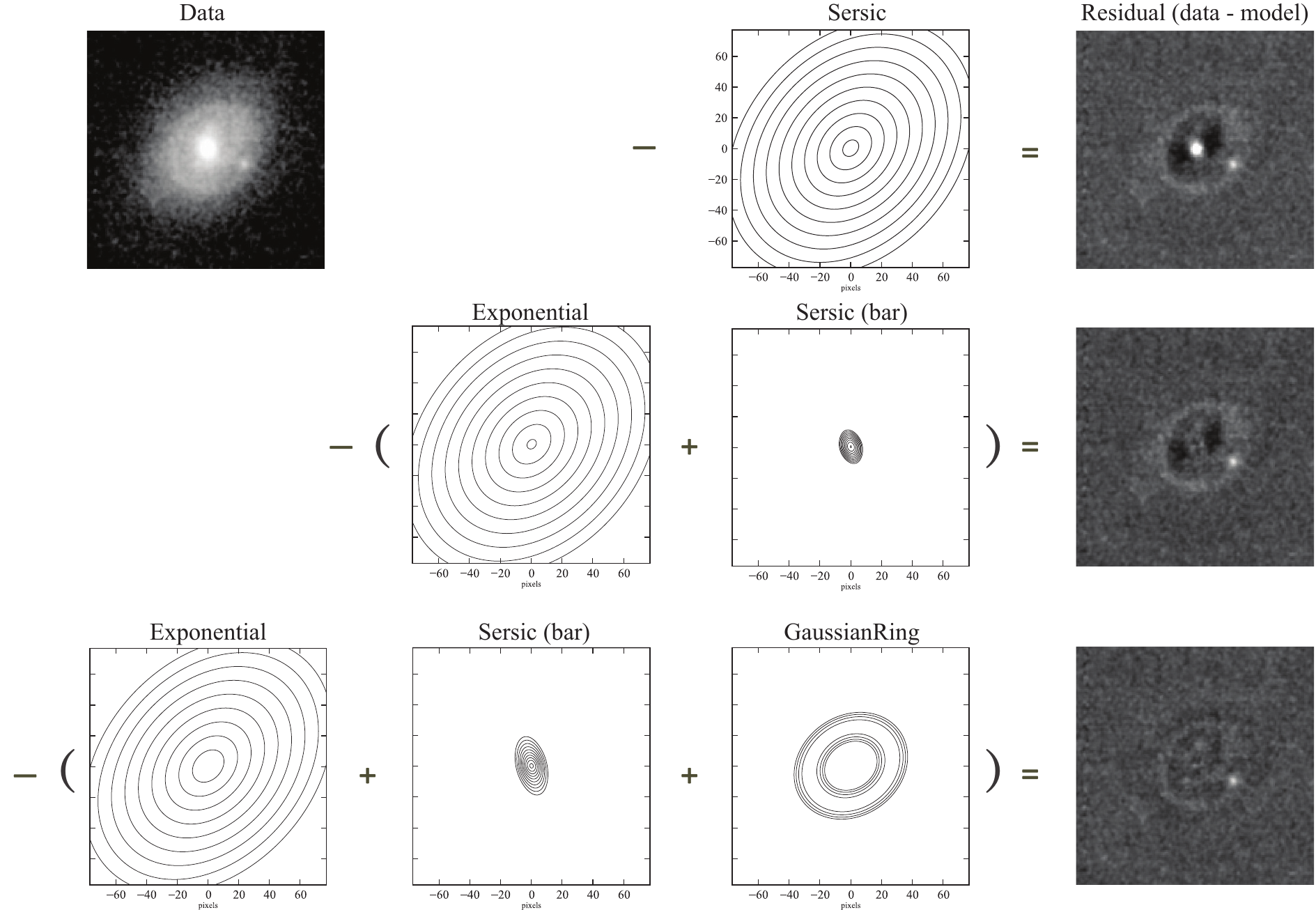}
\end{center}

\caption{Fits of progressively more complex models to a narrow-band
continuum image of the spiral galaxy PGC~35772. Top row: data image
(displayed with log stretch), log-scaled isophote contours
of best-fit Sersic model, residual image (data $-$ model) image,
displayed with linear stretch (1 pixel = 0.238\arcsec). Middle
row: isophote contours of best-fit Exponential + Sersic components,
residual image. Bottom row: isophote contours of best fit Exponential +
Sersic + GaussianRing components, residual image.\label{fig:haggis}}

\end{figure*}

PGC 35772 is a $z = 0.0287$, early/intermediate-type spiral galaxy
(classified as SA0/a by \citealt{rc3} and as Sb by \citealt{fukugita07})
which was observed as part of the H$\alpha$ Galaxy Groups Imaging Survey
(HAGGIS; Kulkarni et al., in prep.) using narrow-band filters on the
Wide Field Imager of the ESO 2.2m telescope. The upper-left panel of
Figure~\ref{fig:haggis} shows the stellar-continuum-filter image
(central wavelength $\approx 659$~nm, slightly blueward of the
redshifted H$\alpha$ line). Particularly notable is a bright stellar
ring, which makes this an interesting test case for including rings in
2D fits of galaxies. Ellipse fits to the image show strong twisting of
the isophotal position angle interior to the ring, suggesting a bar is
also present.

The rest of Figure~\ref{fig:haggis} shows the results three different
fits to the image, each successive fit adding an extra component. These
fits use a $291 \times 281$-pixel cutout of the full WFI image, and were
convolved with a Moffat-function image with FWHM = 0.98\arcsec,
representing the mean PSF (based on Moffat fits to stars in the same
image). The best-fit parameters for each model, determined by minimizing
\pmlr, are listed in Table~\ref{tab:haggis}, along with the uncertainties
estimated from the L-M covariance matrix. Since the fitting times are short, I
also include parameter uncertainties from 500 rounds of bootstrap resampling
(in parentheses, following the L-M uncertainties).

The first fit is uses a single Sersic component; the residuals of this
fit show a clear excess corresponding to the ring, as well as
mis-modeling of the region inside the ring. The fit is improved by
switching to an exponential + S\'ersic model, with the former component
representing the main disk and the latter some combination of the bar +
bulge (if any). This two-component model (middle row of the figure)
produces less extreme residuals;  the best-fitting S\'ersic component is
elongated and misaligned with the exponential component,
so it can be seen to be modeling the bar.

The residuals to this ``disk + bar'' fit are still significant, however,
including the ring itself. To fix this, I include a GaussianRing
component (Section~\ref{sec:gaussian-ring}) in the third fit (bottom row of
Figure~\ref{fig:haggis}). The residuals to \textit{this} fit are better
not just in the ring region, but also inside, indicating that this
three-component model is doing a much better job of modeling the inner
flux of the galaxy (the three-component also has the smallest AIC value of
the three models; see Table~\ref{tab:haggis}).

\begin{deluxetable}{llrll}
\tablecaption{Results of Fitting PGC~35772\label{tab:haggis}}
\tablecolumns{5}
\tablehead{\colhead{Component} & \colhead{Parameter}  & \colhead{Value} & \colhead{$\sigma$} & \colhead{units} \\
\colhead{(1)}    & \colhead{(2)}        & \colhead{(3)}    & \colhead{(4)}   & \colhead{(5)}}
\startdata
\multicolumn{5}{c}{Sersic only (AIC $= 18419$)} \\
Sersic        & PA         & 138.14 & 0.48 (0.34)    & deg \\
              & $\epsilon$ & 0.254  & 0.0038 (0.0024)  &     \\
              & $n$        & 1.041  & 0.0085 (0.027)  &     \\
              & $I_{e}$    & 39.16  & 0.33 (0.52)    & cont.\ flux \\
              & $r_{e}$    & 8.267  & 0.042 (0.034)   & arcsec \\
\\[0.5 mm]
\multicolumn{5}{c}{Exponential + Bar (AIC $= 16569$)} \\
Exponential   & PA         & 137.79 & 0.49 (0.29)    & deg \\
(disk)        & $\epsilon$ & 0.259  & 0.0039 (0.0021)  &     \\
              & $I_{0}$    & 194.58 & 1.22 (0.97)    & cont.\ flux \\  
              & $h$        & 5.17   & 0.025 (0.014)   & arcsec \\
Sersic        & PA         & 16.27  & 3.00 (1.13)    & deg \\
(bar)         & $\epsilon$ & 0.562  & 0.056 (0.025)   &     \\
              & $n$        & 0.897  & 0.282 (0.074)   &     \\
              & $I_{e}$    & 171.82 & 25.32 (6.77)   & cont.\ flux  \\
              & $r_{e}$    & 0.713  & 0.041 (0.021)   & arcsec \\
\\[0.5 mm]
\multicolumn{5}{c}{Exponential + Bar + Ring (AIC $= 14996$)} \\
Exponential   & PA         & 140.70 & 1.25 (0.53)    & deg \\
(disk)        & $\epsilon$ & 0.277  & 0.0098 (0.0053)  &     \\
              & $I_{0}$    & 111.19 & 11.55 (4.05)   & cont.\ flux \\
              & $h$        & 5.74   & 0.18 (0.052)    & arcsec \\
Sersic        & PA         & 7.27   & 2.28 (1.04)    & deg \\
(bar)         & $\epsilon$ & 0.364  & 0.028 (0.092)   &     \\
              & $n$        & 1.14   & 0.114 (0.046)   &     \\
              & $I_{e}$    & 80.78  & 7.25 (2.47)    & cont.\ flux \\
              & $r_{e}$    & 1.42   & 0.080 (0.028)   & arcsec \\
GaussianRing  & PA         & 128.40 & 1.69 (0.69)    & deg \\
(ring)        & $\epsilon$ & 0.258  & 0.013 (0.0053)   &     \\
              & $A$        & 26.90  & 3.22 (1.10)    & cont.\ flux \\
              & $R$        & 5.50   & 0.36 (0.14)    &  arcsec \\
              & $\sigma$   & 3.43   & 0.22 (0.13)    & arcsec \\
\enddata

\tablecomments{Results of fitting narrow-band continuum image of spiral galaxy
PGC~35771 with progressively more complex models (Sersic; Exponential +
Sersic; Exponential + Sersic + GaussianRing).  ``AIC'' = Akaike
Information Criterion values for the fits; lower values imply better
fits. Column 1: Component used in fit. Column 2: Parameter. Column
3: Best-fit value for parameter. Column 4: Uncertainty on parameter
value from L-M covariance matrix; uncertainty from bootstrap resampling is in
parentheses. Column 5: Units (``cont.\ flux'' units are $10^{-18}$ erg
s$^{-1}$ cm$^{-2}$ \AA$^{-1}$ arcsec$^{-2}$).}
\end{deluxetable}

\subsection{IC 5176: Fitting Thin and Thick Disks in an Edge-on Spiral in 2D and 3D} 



IC~5176 is an edge-on Sbc galaxy, included in a ``control'' sample of
non-boxy-bulge galaxies by \citet{chung04} and \citet{bureau06}.
\citet{chung04} noted that both the gas and stellar kinematics were
consistent with an axisymmetric, unbarred disk; \citet{bureau06}
concluded from their $K$-band image that it had a very small bulge and a
``completely featureless outer (single) exponential disk.''  This
suggests an agreeably simple, axisymmetric structure, ideal for an
example of modeling an edge-on galaxy. To minimize the effects of the
central dust lane (visible in optical images of the galaxy), I use a
\textit{Spitzer} IRAC1 (3.6~\micron) image from S4G \citep{sheth10},
retrieved from the \textit{Spitzer} archive. For PSF convolution, I use
an in-flight point response function image for the center of the IRAC1
field,\footnote{\href{http://irsa.ipac.caltech.edu/data/SPITZER/docs/irac/calibrationfiles/}{http://irsa.ipac.caltech.edu/data/SPITZER/docs/irac/calibrationfiles/}} downsampled to the 0.6\arcsec{} pixel scale of the
post-processed archival galaxy image.

\begin{figure}
\begin{center}
\includegraphics[scale=0.45]{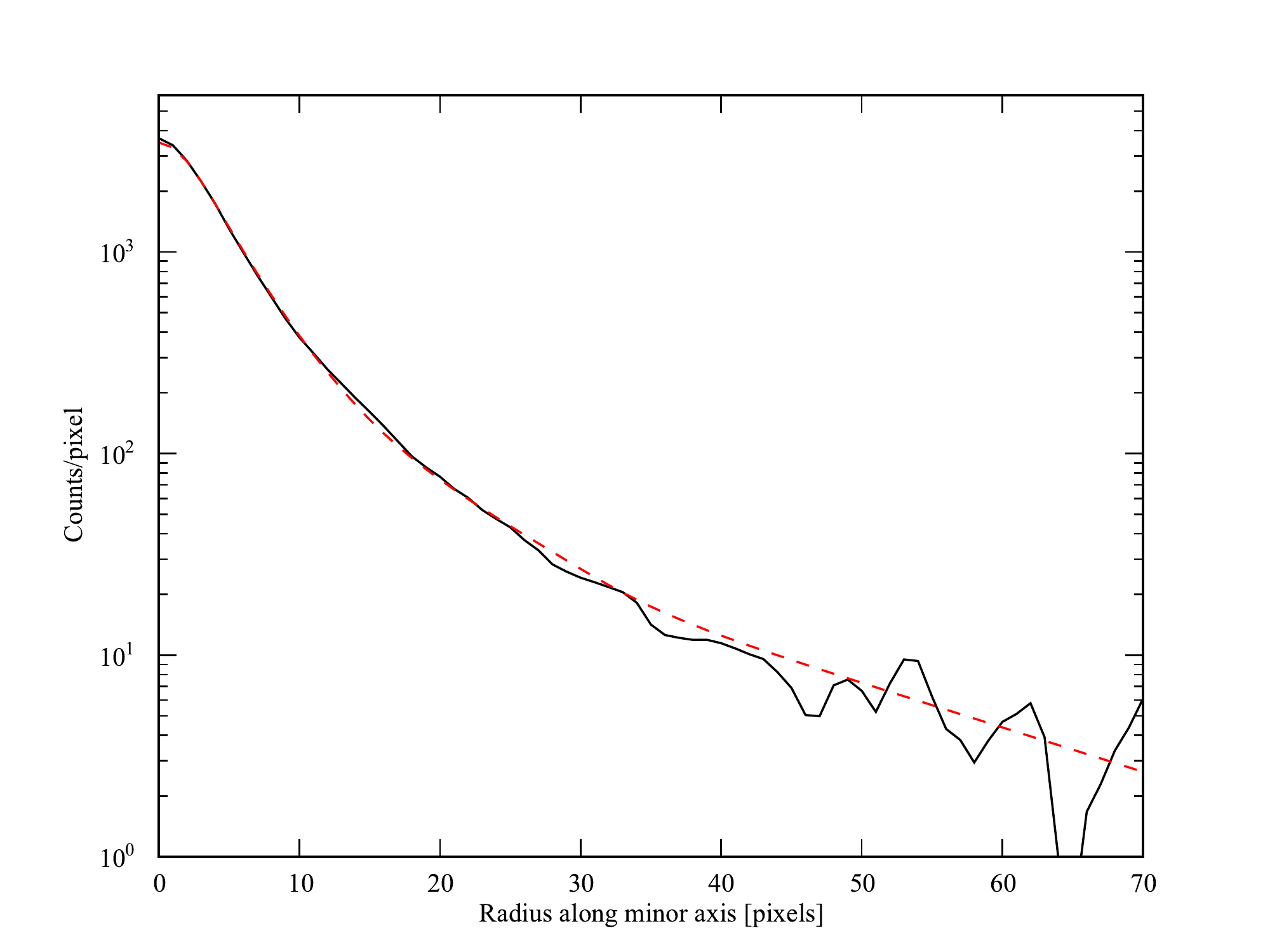}
\end{center}

\caption{Minor-axis profile from \textit{Spitzer} IRAC1 (3.6~\micron)
image of edge-on spiral galaxy IC~5176 (black line), along with corresponding profile from
best-fitting two-disk model (red dashed line).\label{fig:ic5176-prof}}

\end{figure}

Inspection of major-axis and minor-axis profiles from the IRAC1 image
(Figure~\ref{fig:ic5176-prof}) suggests the presence of both thin and
thick disk components;  the $K$-band image of \citet{bureau06} was
probably not deep enough for this to be seen. The major axis profile
and the image both suggest a rather round, central excess, consistent
with the small bulge identified by \nocite{bureau06}Bureau et al. 

Consequently, I fit the image using a combination of two
exponential-disk models, plus a central S\'ersic component for the
bulge. The fast way to fit such a galaxy with \imfit{} is to assume that
the galaxy is perfectly edge-on and use the 2D analytic EdgeOnDisk
functions (Section~\ref{sec:edgeondisk}) for the thin and thick disk
components. Table~\ref{tab:ic5176} shows the results of this fit. The
dominant EdgeOnDisk component, which can be thought of as the ``thin
disk'', has a nearly sech vertical profile and a scale height of
$2.0\arcsec \approx 260$ pc \citep[assuming a distance of 26.4
Mpc;][]{tully09}. The second EdgeOnDisk, with a more exponential-like
vertical profile and a scale height of 1.4 kpc, is then the ``thick
disk'' component; it has a radial scale length $\sim 2.9$ times that of
the thin disk.

The central S\'ersic component of this model contributes 1.8\% of the
total flux, while the thin and thick disks account for 70.5\% and
27.7\%, respectively. The thick/thin-disk luminosity ratio of 0.39 is
consistent with the recent study of thick and thin disks by
\citet{comeron11b}: using their two assumed sets of relative
mass-to-light ratios gives a mass ratio $M_{\rm thick}/M_{\rm thin} =
0.47$ or 0.94, which places IC~5176 in the middle of the distribution
for galaxies with similar rotation velocities (see their Fig.~13).



%
%

\begin{deluxetable}{llrrl}
\tablecaption{Results of Fitting IC 5176\label{tab:ic5176}}
\tablecolumns{5}
\tablehead{\colhead{Component} & \colhead{Parameter}  & \colhead{Value} & \colhead{$\sigma$} & \colhead{units} \\
\colhead{(1)}    & \colhead{(2)}        & \colhead{(3)}    & \colhead{(4)}   & \colhead{(5)}}
\startdata
\multicolumn{5}{c}{Fit with 2D Disks (AIC $= 182129$)} \\
Sersic        & PA         & 149.7  & 0.0049 & deg \\
(bulge)       & $\epsilon$ & 0.206  & 0.014   &     \\
              & $n$        & 0.667  & 0.033   &     \\
              & $\mu_{e}$  & 12.90  & 0.000  & mag~arcsec$^{-2}$ \\
              & $r_{e}$    & 1.48   & 0.019    & arcsec \\
EdgeOnDisk    & PA         & 149.7  & 0.0049   & deg \\
(thin disk)   & $\mu_{0}$  & 11.829  & 0.0008   & mag~arcsec$^{-2}$   \\
              & $h$        & 14.17  & 0.012   & arcsec \\
              & $n$        & 2.607  & 0.025   &  \\
              & $z_{0}$    & 2.01   & 0.0044    & arcsec \\
EdgeOnDisk    & PA         & 151.3  & 0.019   & deg \\
(thick disk)  & $\mu_{0}$  & 15.557  & 0.0057   & mag~arcsec$^{-2}$    \\
              & $h$        & 40.97  & 0.011   & arcsec \\
              & $n$        & 9.89   & 0.700    &  \\
              & $z_{0}$    & 10.88  & 0.036   & arcsec \\
\\[0.5 mm]
\multicolumn{5}{c}{Fit with 3D Disks (AIC $= 179824$)} \\
Sersic        & PA         & 168.71 & 9.73    & deg \\
(bulge)       & $\epsilon$ & 0.046  & 0.016   &     \\
              & $n$        & 0.762  & 0.033   &     \\
              & $\mu_{e}$  & 13.10  & 0.023  & mag~arcsec$^{-2}$  \\
              & $r_{e}$    & 1.46   & 0.019    & arcsec \\
ExponentialDisk3D  & PA    & 149.73 & 0.001   & deg \\
(thin disk)   & $i$        & 87.21  & 0.015    & deg \\
              & $\mu_{0}$  & 11.475  & 0.0010   & mag~arcsec$^{-2}$    \\
              & $h$        & 14.44  & 0.011   & arcsec \\
              & $n$        & 50     & ---      &  \\
              & $z_{0}$    & 2.04   & 0.004    & arcsec \\
ExponentialDisk3D  & PA    & 151.43 & 0.019    & deg \\
(thick disk)  & $i$        & 89.40  & 0.126    & deg \\
              & $\mu_{0}$  & 15.604  & 0.0040   & mag~arcsec$^{-2}$    \\
              & $h$        & 42.07  & 0.115   & arcsec \\
              & $n$        & 50     & ---      &  \\
              & $z_{0}$    & 11.74  & 0.038   & arcsec \\
\enddata

\tablecomments{Results of fitting \textit{Spitzer} IRAC1 (3.6~\micron)
image of the edge-on spiral IC~5176. The first fit uses analytic 2D
EdgeOnDisk components (exponential disk seen at $i = 90\arcdeg$); the
second fit uses line-of-sight integration through ExponentialDisk3D
components (3D luminosity-density models), for which the inclination $i$
is a free parameter. Size parameters have been converted from pixels to
arc seconds; ``AIC'' = Akaike Information Criterion values for the two
fits. Column 1: Component used in fit. Column 2: Parameter. Column
3: Best-fit value for parameter. Column 4: Uncertainty on parameter
value from L-M covariance matrix. Column 5: Units (surface-brightness
parameters have been converted from counts/pixel to 3.6$\mu$m AB mag
arcsec$^{-2}$; the $\mu_0$ values for the disk components are equivalent
integrated face-on central surface brightnesses). }

\end{deluxetable}

A slower but more general approach is to use the EdgeOnDisk3D function
(Section~\ref{sec:expdisk3d}) for both components, which allows for
arbitrary inclinations. The cost is in the time taken for the fit: $\sim
29$ minutes, versus a mere 3m20s for the analytic 2D approach. Using the
EdgeOnDisk3D functions \textit{does} give what is formally a better
model of the data than the analytic 2D-component fit, with $\Delta$AIC
$\approx 2305$, though most of the parameter values -- in particular,
the radial and vertical scale lengths -- are almost identical to
previous fit. The only notable changes are the S\'ersic component
becoming rounder (with a different and probably not very well-defined
position angle) and the vertical profiles of both disk components
becoming pure exponentials (the values of $n$ in Table~\ref{tab:ic5176}
are imposed limits). The relative contributions of the three components
are essentially unchanged: 1.8\% of the flux from the S\'ersic component
and 71.7\% and 26.5\% from the thin and thick disks, respectively. The
best-fitting model converges to $i \approx 90\arcdeg$ for the outer
(thick) disk component, but does find $i = 87.2\arcdeg$ for the
thin-disk component.  

\begin{figure*}
\begin{center}
\includegraphics[scale=1.0]{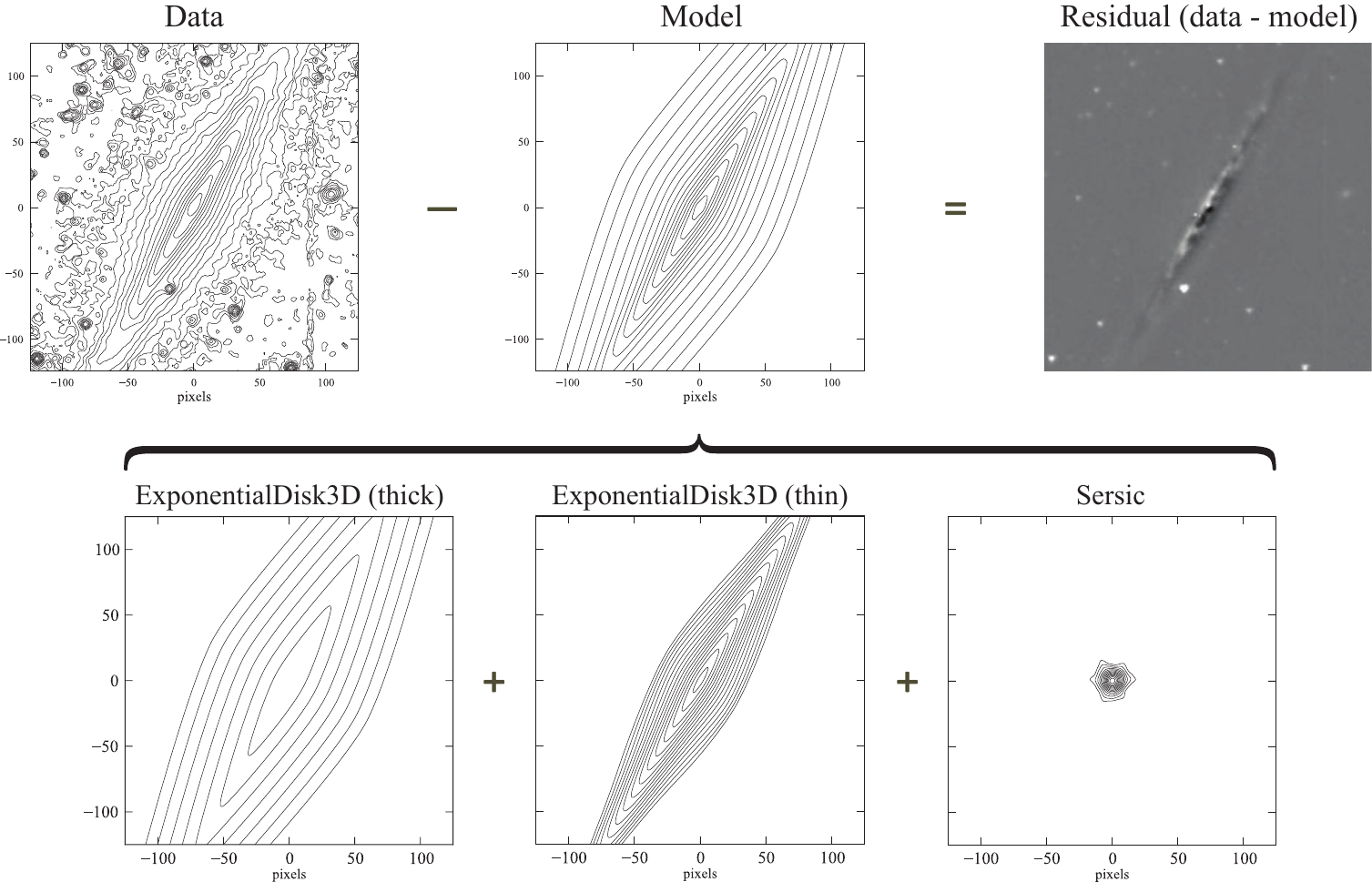}
\end{center}

\caption{Top row, left: Logarithmically scaled isophotes of the
\textit{Spitzer} IRAC1 (3.6~\micron) image of edge-on spiral galaxy IC~5176, smoothed with a
5-pixel-wide median filter (1 pixel = 0.6\arcsec). Top row, middle:
Best-fitting, PSF-convolved model (see bottom row). Top row, right:
residual image (data $-$ model), displayed with linear stretch. Bottom
row: log-scaled isophotes showing PSF-convolved components making up the
best-fitting model, consisting of two ExponentialDisk3D components and a
Sersic component. All isophote plots use the same logarithmic scaling.\label{fig:ic5176}}

\end{figure*}

The reality is that the combination of low spatial resolution of the
IRAC1 image and the presence of residual structure in the disk midplane
(probably due to a combination of spiral arms, star formation, and dust)
means that we cannot constrain the vertical structure of the disk(s)
very well. A vertical profile which is best fit with a sech function when
the disk is assumed to be perfectly edge-on can also be fit with a
vertical exponential function, if the disk is tilted slightly from edge-on. The
low spatial resolution also means that the central bulge is not well
constrained, either; the half-light radius of the S\'ersic component
from either fit is $\sim 2.5$ pixels and thus barely larger than the
seeing.

%
%
%
%
 
%
%

\section{Potential Biases in Fitting Galaxy Images: \chisquare{} Versus Poisson Maximum-Likelihood Fits}
\label{sec:biases} 

In Section~\ref{sec:statistics}, I discussed two different
practical approaches to fitting images from a statistical point of view:
the standard, Gaussian-based \chisquare{} statistic and Poisson-based
MLE statistics ($C$ and \pmlr). The \chisquare{} approach
can be further subdivided into the common method of using data values to
estimate the per-pixel errors (\chisquaredata) and the alternate method
of using values from the model (\chisquaremodel). Outside of certain
low-S/N contexts (e.g., fitting X-ray and gamma-ray data), \chisquare{}
minimization is pretty much the default. Even in the case of
low S/N, when the Gaussian approximation to Poisson statistics -- which
motivates the \chisquare{} approach -- might start to become invalid,
one might imagine that the presence of Gaussian read noise in CCD
detectors could make this a non-issue. Is there any reason for using
Poisson-likelihood approaches outside of very-low-count, zero-read-noise
regimes?

\citet{humphrey09} used a combination of analytical approximations and
fits of models to artificial data to show how \chisquare{} fits
(using data-based or model-based errors) can lead to biased parameter
estimation, even for surprisingly high S/N ratios; these biases were
essentially absent when Poisson MLE was used. (\citealt{humphrey09}
used $C$ for their analysis, but minimizing \pmlr{} would yield the same
fits, as noted in Section~\ref{sec:poisson}.) A fuller discussion of
these issues in the context of fitting X-ray data can be found in that
paper, and references therein \citep[e.g.][]{nousek89}. In this section,
I focus on the typical optical imaging problem of fitting galaxy images
with simple 2D functions and use the flexibility of \imfit{} to explore
how fitting Poisson (or Poisson + Gaussian) data with different
assumptions can bias the resulting fitted parameter values.

\subsection{Fitting Simple Model Galaxy Images}\label{sec:model-images} 

As a characteristic example, I consider a model galaxy described by a
2D S\'ersic function with $n = 3.0$, $r_{e} = 20$ pixels, and an
ellipticity of 0.5. This model is realized in three count-level regimes:
a ``low-S/N'' case with a sky background level of 20 counts/pixel and
model intensity at the half-light radius $I_{e} = 50$ counts/pixel; a
``medium-S/N'' version which is equivalent to an exposure time (or
telescope aperture) five times larger (background level = 100, $I_{e} =
250$); and a ``high-S/N'' version with total counts equal to 25 times
the low-S/N version (background level = 500, $I_{e} = 1250$). These
values are chosen partly to test the question of how rapidly the
Gaussian approximation to Poisson statistics becomes appropriate: 20
counts/pixel is often given as a reasonable lower limit for using this
approximation \citep[e.g.,][]{cash79}, while for 500
counts/pixel the Gaussian approximation should be effectively
indistinguishable from true Poisson statistics.


\begin{figure*}
\begin{center}
\includegraphics[scale=0.98]{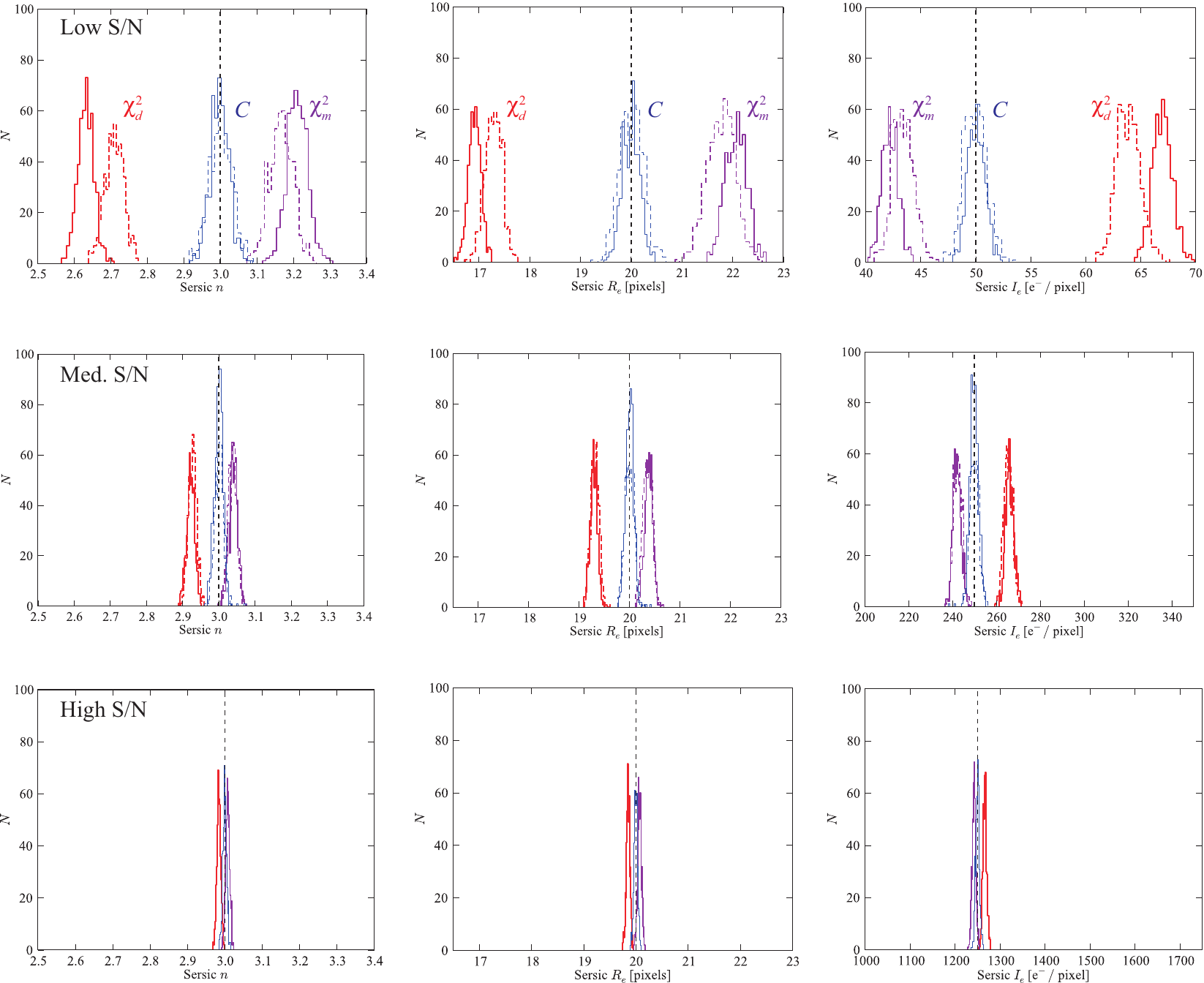}
\end{center}

\caption{Distribution of best-fit parameters from fits to 500
realizations of an artificial galaxy image with an elliptical S\'ersic
component + sky background and pure Poisson noise (solid histograms), or
Poisson noise + Gaussian read noise with $\sigma = 5$ e$^{-}$ (dashed
histograms). Fits used data-based \chisquare{} minimization
(\chisquaredata, red histograms), model-based \chisquare{} minimization
(\chisquaremodel, magenta), or Poisson MLE minimization (using $C$, blue);
vertical dashed gray lines indicate parameter values of the original
model. \textit{Upper row:} Results for low-S/N images (sky background =
20 e$^{-}$/pixel, S\'ersic model $I_{e} = 50$ e$^{-}$/pixel).
\textit{Middle row:} Results for medium-S/N images (background = 100
e$^{-}$/pixel, $I_{e} = 250$ e$^{-}$/pixel). \textit{Bottom row:}
Results for high-S/N images (background = 500 e$^{-}$/pixel, $I_{e} =
1250$ e$^{-}$/pixel); the additional histograms for fits to images with
Gaussian read noise are in this case essentially indistinguishable from
the pure-Poisson-noise histograms and are not plotted. Using
\chisquaredata{} minimization systematically underestimates (S\'ersic
$n$ and $r_{e}$) or overestimates ($I_{e}$) the parameters, while using
\chisquaremodel{} minimization produces smaller biases in the opposite
directions. These biases diminish as the counts/pixel get larger. The
presence of Gaussian read noise reduces the \chisquare{} bias in the
low-S/N regime, but does not eliminate it. In all cases, minimization 
of the Poisson MLE statistic $C$ is bias-free. \label{fig:sersic-histograms}}

\end{figure*}

The images were created using code written in Python. The
first stage was generating a noiseless $150 \times 150$-pixel reference
image (including subpixel integration, but not PSF convolution). This
was then used as the source for generating 500 ``observed'' images of
the same size, using Poisson statistics: for each pixel, the value in
the reference image was taken as the mean $m$ for a Poisson process
(Equation~\ref{eq:poisson}), and actual counts were (pseudo)randomly
generating using code in the Numpy package
(\texttt{numpy.random.poisson}).\footnote{\texttt{http://www.numpy.org}}
For simplicity, the gain was set to 1, so 1 count = 1 photoelectron.

The resulting images were then fit with \imfit{} three times, always
using the Nelder-Mead simplex method as the minimization algorithm. The
first two fits used \chisquare{} statistics, either the data-based
\chisquaredata{} or the model-based \chisquaremodel{} approach, with
read noise set to 0; the third fit minimized $C$.
(Essentially identical fits are obtained when minimizing \pmlr{}
instead of $C$.) The fitted model consisted of a single 2D S\'ersic
function with very broad parameter limits and the same starting
parameter values for all fits (with the initial $I_{e}$ value scaled by
5 for the medium-S/N images and by 25 for the high-S/N images), along
with a fixed FlatSky component for the background.

Figure~\ref{fig:sersic-histograms} shows the distribution of best-fit
parameters for fits to all 500 individual images in each S/N regime,
with thick red histograms for the \chisquaredata{} fits, thinner magenta
histograms for the \chisquaremodel{} fits, and thin blue histograms for
the Poisson MLE ($C$) fits, along with the true parameter values of the
original model as vertical dashed gray lines.

\begin{figure*}
\begin{center}
\includegraphics[scale=0.88]{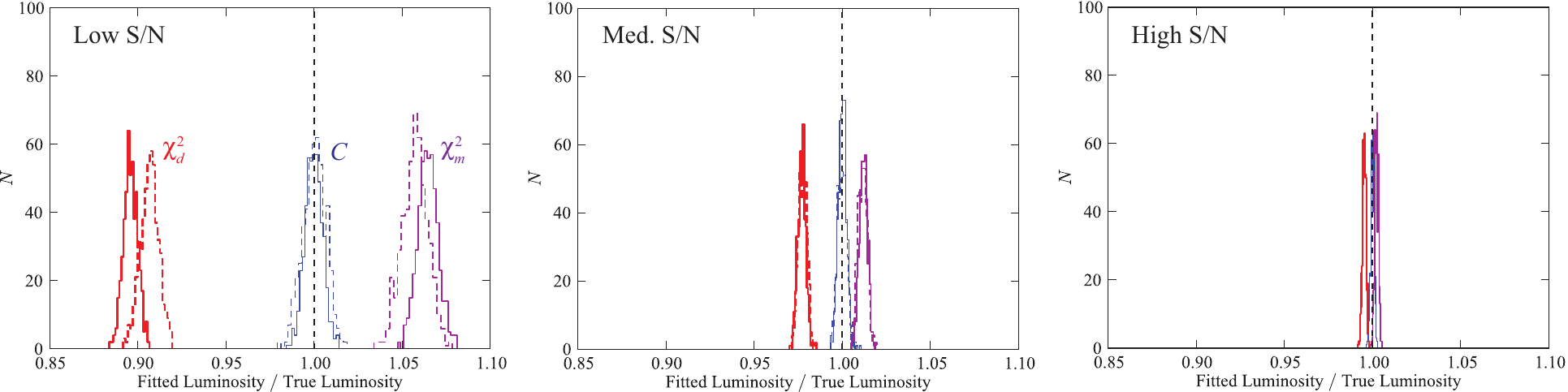}
\end{center}

\caption{As for Figure~\ref{fig:sersic-histograms}, but now showing the
distribution of the estimated (S\'ersic) galaxy luminosity (relative to
the true luminosity) from the fits to the model images. In the low-S/N
case (left panel), the data-based \chisquaredata{} fits (red)
underestimate the true luminosity by 10.4\% (9.3\% when read noise is
present, dashed red histogram), while the model-based \chisquaremodel{}
fits (magenta) overestimate it by 6.5\% (5.8\% for the read-noise case);
the Poisson MLE fits ($C$, blue) are unbiased. These biases diminish in
the medium-S/N regime (middle panel) and high-S/N regime (right panel).
\label{fig:lum-histograms}}

\end{figure*}

A clear bias for the \chisquare{} approaches can be seen in the fits to
the  low-S/N images (top panels of Figure~\ref{fig:sersic-histograms}).
For the \chisquaredata{} approach, the fitted values of $n$ and $r_{e}$
are too small: the average value of $n$ is 12.3\% low, while the average
value of $r_{e}$ is 15.4\% too small. The fitted values of $I_{e}$, on
the other hand, are on average 34\% too large. As can be seen from the
figure, these biases are significantly larger than the spread of values
from the individual fits.  The overall effect also biases the total flux
for the S\'ersic component, which is underestimated by 10.4\% when using
the mean parameters of the \chisquaredata{} fit; see
Figure~\ref{fig:lum-histograms}. The (model-based) \chisquaremodel{}
approach also produces biases, though these are smaller and are in the
opposite sense from the \chisquaredata{} biases: $n$ and $r_{e}$ are
overestimated on average by 7.0\% and 10.4\%, respectively, while
$I_{e}$ is 15.6\% too small; the total flux is overestimated by 6.5\%.
Finally, the fits using $C$ are \textit{unbiased}: the
histograms straddle the input model values, and the mean values from the
fits are all $< 0.1$\% different from the true values. The other
parameters of the fits -- galaxy center, position angle, ellipticity --
do not show any systematic differences, except for a very slight
tendency of the ellipticity to be biased high with the \chisquaredata{}
fit, but only at the $\sim 0.5$\% level. For the parameters which show
biases in the \chisquare{} fits, the trends are exactly as suggested by
\citet{humphrey09}, including the fact that the \chisquaremodel{} biases
are smaller and have the opposite sign from the \chisquaredata{} biases.


In the medium-S/N case (middle panels of the same figure), the bias in
the \chisquaredata{} and \chisquaremodel{} fits is clearly reduced: for
the \chisquaredata{} fits, $n$ and $r_{e}$ are on average only 2.6\% and
3.5\% too small, while $I_{e}$ is on average 6.4\% too high (in the
\chisquaremodel{} fits, the deviations are 1.3\% and 1.9\% too large and
3.2\% too small, respectively) -- though the bias is clearly still present,
and in the same pattern. The biases in total flux are smaller, too:
2.3\% low and 1.2\% high for the data-based and model-based \chisquare{}
fits, respectively (Figure~\ref{fig:lum-histograms}). These biases are
even smaller in the high-S/N case: e.g., in the \chisquaredata{} case,
$n$ and $r_{e}$ are 0.54\% and 0.74\% too small, while $I_{e}$ is 1.3\%
too high. In both S/N regimes, the Poisson MLE fits remain unbiased.


What is the effect of adding (Gaussian) read noise to the images? To
investigate this, additional sets of images were prepared as before,
except that the value from the Poisson process was further modulated by
adding a Gaussian with mean = 0 and width $\sigma = 5$ e$^{-1}$. (This
value was chosen as a representative read noise for typical modern CCDs;
it is also roughly equal to the dispersion of the Gaussian approximation
to the \textit{Poisson} noise of the background in the low-S/N limit --
i.e., $\sigma_{\rm sky} \approx \sqrt{20}$.)

The fits were done as before, with the read noise properly included in
the \chisquare{} fitting; the histograms of the resulting fits are shown
in Figures~\ref{fig:sersic-histograms} and \ref{fig:lum-histograms} with
dashed lines. What is clear from the figure is that while the addition
of a Gaussian noise term reduces the bias in the \chisquare{} fits
slightly in the low-S/N regime, the bias is still present.
Even though the Poisson MLE approach is no longer formally correct when
Gaussian noise is present, the $C$ fits remain unbiased in
the presence of moderate read noise.


\begin{figure*}
\begin{center}
\includegraphics[scale=0.9]{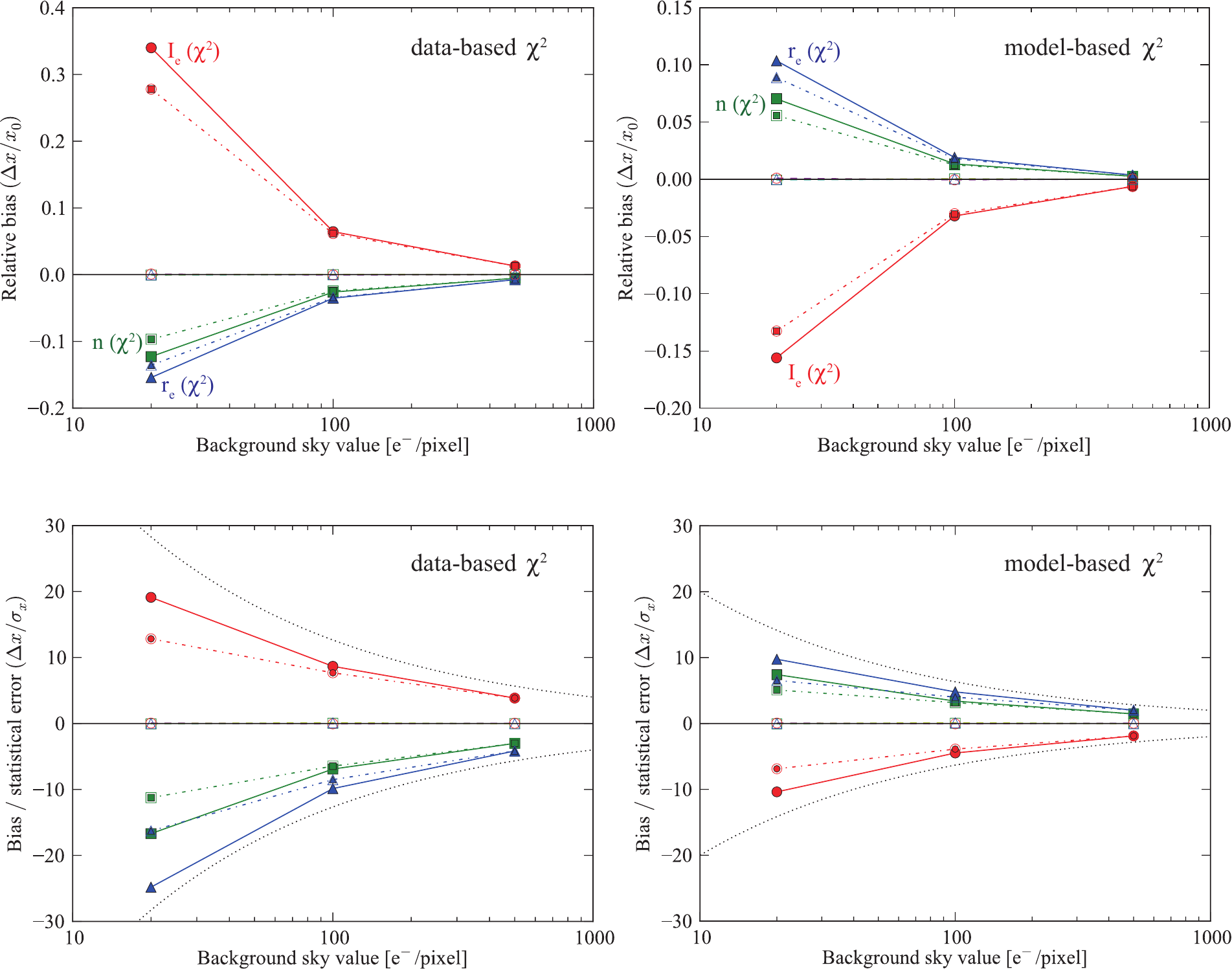}
\end{center}

\caption{Bias in fitted S\'ersic parameters, for the low-, medium-, and
high-S/N model images (see Figure~\ref{fig:sersic-histograms}). The bias
is plotted versus the background level of the corresponding images.
Solid symbols and lines are for parameter values from \chisquare{} fits
to pure-Poisson images (green = $n$, red = $I_{e}$, blue = $r_{e}$),
while semi-filled symbols and dot-dashed lines are from \chisquare{}
fits to images with added read noise. Hollow symbols and dashed lines are
from Poisson MLE fits, which show essentially no bias. \textit{Top
panels}: Fractional bias $(\bar{x} - x_{0})/x_{0}$, where $\bar{x}$ is
the mean measured parameter value from fits to 500 images and $x_{0}$ is
the original model value; the left and right panels show
\chisquaredata{} and \chisquaremodel{} fits, respectively.
\textit{Bottom panels:} Same, but now showing bias relative to
statistical error $(\bar{x} - x_{0})/\sigma_{x}$, where $\sigma_{x}$ is
the nominal statistical error from the fit. The upper and lower dotted
curves in each panel show the predicted bias from \citet{humphrey09},
which is based purely on total counts and number of pixels in the
images (Equations~\ref{eq:bias-estimate1} and \ref{eq:bias-estimate2}). \label{fig:sersic-deviations}}

\end{figure*}


\begin{figure*}
\begin{center}
\includegraphics[scale=0.9]{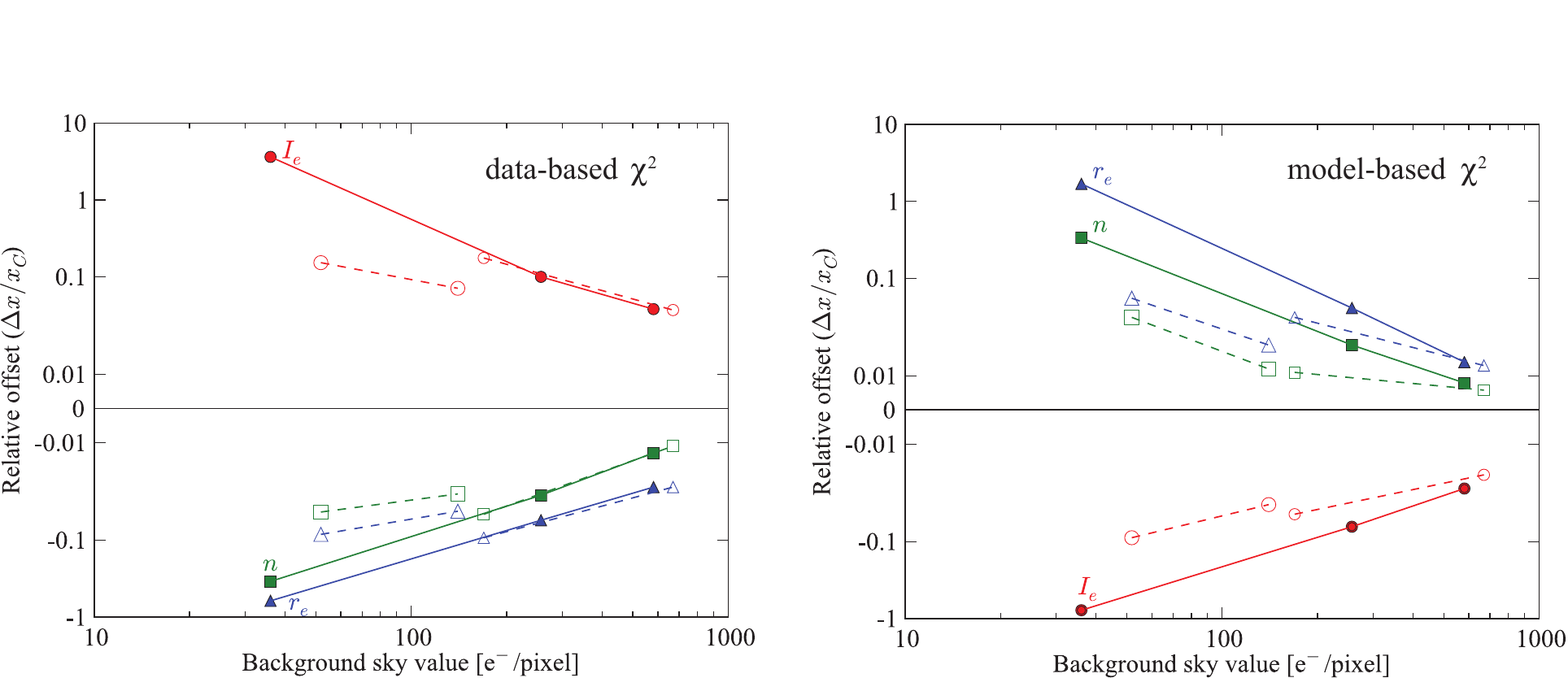}
\end{center}

\caption{As for the top panels of Figure~\ref{fig:sersic-deviations},
but now showing relative differences in fitted S\'ersic parameters
between \chisquare{} fits and Poisson MLE fits to images of three
real elliptical galaxies. Shown is $(x -
x_{C})/x_{C}$, where $x$ is the $n$, $r_{e}$, or
$I_{e}$ value from a \chisquare{} fit and $x_{C}$ is the value from
a Poisson MLE fit to the same image, plotted against mean sky
background for the image. \textit{Left panel:} Data-based \chisquare{}
fits. \textit{Right panel:} Model-based \chisquare{} fits. Solid points
are from fits to (in order of increasing background level) SDSS $u$,
$g$, and $r$ images of NGC~5831; small hollow symbols are from fits to
15s and 60s INT-WFC $r$-band images of NGC~4697, while larger hollow
symbols are from fits to 15s and 40s INT-PFCU $V$-band images of
NGC~3379. Green squares, blue triangles, and red circles indicate
S\'ersic $n$, $r_{e}$, and $I_{e}$, respectively; solid lines connect
results for the same galaxy with different exposure levels.
\label{fig:real-deviations}}

\end{figure*}

\subsection{Quantifying the Bias} 

How large is the bias produced by \chisquare{} fits? \citet{humphrey09}
suggested that the absolute or relative size of the bias might not be as
important as the size of the bias relative to the nominal statistical
errors of the fits. There are, in principle, three different ways of
estimating these errors: from the distribution of the fitted values for
all 500 images (similar to what was done by Humphrey et al.\ for their
examples); from the mean of individual-fit error estimates produced by
using the L-M algorithm; and from the mean of individual-fit error
estimates produced by bootstrap resampling. For this simple model, all
three approaches produce very similar values. For example, fitting the
images in \chisquaredata{} mode with the L-M algorithm produces
estimated dispersions within $\sim 10$\% of the dispersion of values
from the individual \chisquaredata{} fits; the latter are in turn very
similar to the dispersion of the individual $C$ fits (as is evident from
the similar histogram widths in Figure~\ref{fig:sersic-histograms}). The
errors estimated from bootstrap resampling also agree to within $\sim
10$\% of the other estimates; see Figure~\ref{fig:bootstrap} for a
comparison of bootstrap and L-M error estimates for a fit to a single
low-S/N image.

Figure~\ref{fig:sersic-deviations} shows the biases for the
\chisquaredata, \chisquaremodel, and Poisson MLE fits, plotted
against the background value for the different S/N regimes: the top
panels show the deviations relative to the true parameter values, while
the bottom panels shows the deviations in units of the statistical
errors (using the standard deviation of the 500 fitted values). The left
and right panels show the cases for \chisquaredata{} and
\chisquaremodel{} fits, respectively, with the Poisson MLE fits
shown in each panel for reference. In all cases, there is a clear trend
of the \chisquare{} biases becoming smaller as the overall exposure
level (represented by the mean background level) increases,
asymptotically approaching the zero-bias case exhibited by the
Poisson MLE fits.

\citet{humphrey09} derived an estimate for the bias (relative to the
statistical error) that would result from fitting pure-Poisson data
using the \chisquaredata{} statistic, based on the total number of
counts $N_{c}$ and the total number of bins $N_{\rm bins}$ (i.e., the
total number of fitted pixels): 
\begin{equation}\label{eq:bias-estimate1}
f_{b} (\chisquaredata) \; = \; \frac{|x_{0} - \bar{x}|}{\sigma_{x}} \; \sim \; 
\frac{N_{\rm bins}}{\sqrt{N_{c}}},
\end{equation}
where $x_{0}$ is the true value, $\bar{x}$ is the mean fitted value, and 
$\sigma_{x}$ is the statistical error on the parameter value.
They did the same for the \chisquaremodel{} approach and found
\begin{equation}\label{eq:bias-estimate2}
f_{b} (\chisquaremodel)  \; \sim \; 
-0.5 \; \frac{N_{\rm bins}}{\sqrt{N_{c}}}.
\end{equation}
The estimates derived from these equations are plotted as dotted lines in
Figure~\ref{fig:sersic-deviations}. Although the actual biases are
systematically smaller than the predictions, the overall agreement is
rather good.

%

\subsection{Biases in Fitting Images of Real Galaxies}\label{ref:biases-real} 

Is there any evidence that the \chisquare{}-bias effect is significant when
fitting images of real galaxies? Figure~\ref{fig:real-deviations} shows
the differences seen when fitting single S\'ersic functions to images of
three elliptical galaxies. In the first case, I fit $996 \times
1121$-pixel cutouts from SDSS $u$, $g$, and $r$ images of NGC~5831; these
images correspond to successively higher counts per pixel in both
background and galaxy. (The cutouts, as well as the mask images, were
shifted in $x$ and $y$ to correct for pointing offsets between the
different images.) Although color gradients may produce (genuinely)
different fits for the different images, these should be small for an
early-type galaxy like NGC~5831; more importantly, the bias estimates I
calculate (see next paragraph) are between the \chisquare{} and
Poisson MLE fits for each individual band. In the second and third
cases I fit same-filter images with different exposure times: $1801
\times 1701$-pixel cutouts from short (15s) and long (60s) $r$-band
exposures of NGC~4697, obtained with the Isaac Newton Telescope's Wide
Field Camera on 2004 March 17, and 15s and 40s $V$-band exposures of
NGC~3379, obtained with the Prime Focus Camera Unit of the INT on 1994
March 14 (image size = $1243 \times 1152$ pixels). All images were fit with
a single S\'ersic function, convolved with an appropriate Moffat
PSF image (based on measurements of unsaturated stars in each image).
All fits were done with \chisquaredata, \chisquaremodel, and
Poisson MLE ($C$) minimization; the \chisquare{} fits included appropriate
read-noise contributions (5.8 e$^{-}$ and 4.5 e$^{-}$ for the WFC and
PFCU images, respectively).

Unlike the case for the model images in the preceding section, the
``correct'' S\'ersic model for these galaxies is unknown (as is, for
that matter, the true sky background). Thus,
Figure~\ref{fig:real-deviations} shows the differences between the best
\chisquare{}-fit parameters and the parameters from the Poisson MLE
fits, relative to the value of the latter, instead of the difference
between all three and the (unknown) ``true'' solution. The trends are
nonetheless very similar to the model-image case (compare
Figure~\ref{fig:real-deviations} with the top panels of
Figure~\ref{fig:sersic-deviations}): values of $n$ and $r_{e}$ from the
\chisquaredata{} fits are smaller, and values of $I_{e}$ are larger,
than the corresponding values from the Poisson MLE fits, and the
offsets are reversed when \chisquaremodel{} fitting is done. Although
there is some scatter, the tendency of \chisquaredata{} offsets to be
larger than \chisquaremodel{} offsets is present as well: in fact, the
average ratio of the former to the latter is 1.99 (median = 1.65), which
is strikingly close to the ratio of 2 predicted by \citet{humphrey09}.
Even the fact that the $r_e$ offsets are always larger than the $n$
offsets replicates the pattern from the fits to artificial-galaxy
images. In addition, the offsets between the \chisquare-fit values and
the Poisson MLE fits diminish as the count rate increases, as in the
model-image case. If we make the plausible assumption that the higher
S/N images are more likely to yield accurate estimates of the true
galaxy parameters (to the extent that the galaxies \textit{can} be
approximated by a simple elliptical S\'ersic function), then the
convergence of estimated parameter values in the high-count regime
strongly suggests that the Poisson MLE approach is the least biased
in all regimes.







Of course, for many typical optical and near-IR imaging situations, the
count rates even in the sky background are high enough that differences
between \chisquare{} and Poisson MLE fits can probably be ignored.
For example, typical backgrounds in SDSS $g$, $r$, $i$, and $z$ images
range between $\sim 60$ and 200 ADU/pixel, or $\sim 300$--1000
photoelectrons/pixel.\footnote{Based on measurements of $\sim 25$ SDSS
DR7 fields.} Only for $u$-band images does the background level become
low enough ($\sim 30$--150 photoelectrons/pixel) for the
\chisquare{}-fit bias to become a significant issue.


\subsection{The Origins of the Bias} 

A qualitative explanation for the \chisquaredata{} bias is
relatively straightforward. (A more precise mathematical derivation can
be found in, e.g., \citealt{humphrey09}.) In the low-count regime,
pixels with downward fluctuations from the true model will have
significantly lower $\sigma_{i}$ values than pixels with similar-sized
upward fluctuations; since the weighting for each pixel in the fit is
proportional to $1/\sigma_{i}^{2}$, the downward fluctuations will have
more weight, and so the best-fitting model will be biased downward.

The \chisquaremodel{} bias is slightly more complicated. Here, one has
to consider the effects of different possible models being fitted, because the
$\sigma_{i}$ values are determined by the \textit{model}
values, not the data values. In the low-count regime, a model with
slightly higher flux than the true model will have higher
$\sigma_{i}$ values, which will in turn \textit{lower} the total \chisquare{}. A
model with lower flux will have smaller $\sigma_{i}$ values, which will
increase the total \chisquare. The overall effect will thus be to bias
the best-fitting model upward.

\begin{figure*}
\begin{center}
\includegraphics[scale=0.95]{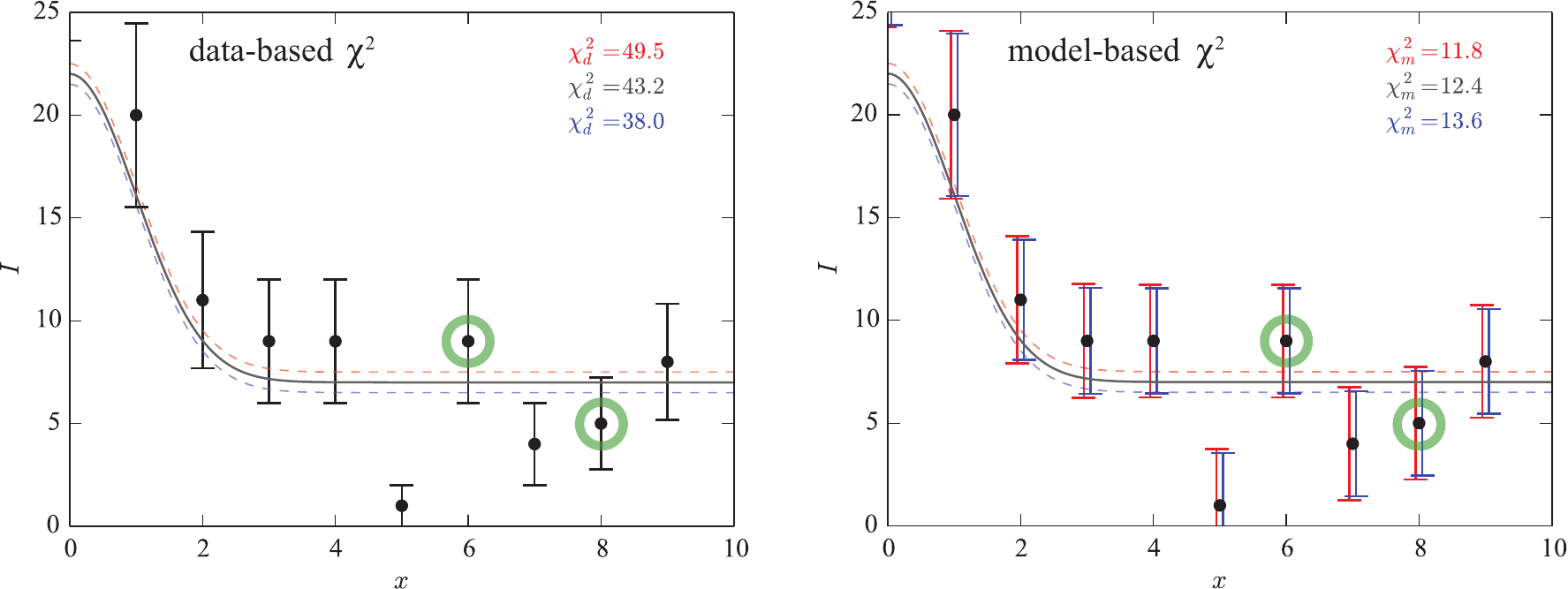}
\end{center}

\caption{A simplified picture of the origin of the \chisquaredata{} and
\chisquaremodel{} biases. Both panels shows the same set of data points
$I(x)$ with Poisson noise generated from a model consisting of a
Gaussian plus a constant background (solid gray line). \textit{Left
panel}: Error bars show 1-$\sigma$ Gaussian uncertainties according to
the $\chisquaredata$ approach ($\sigma = \sqrt{I}$). The data
points at $x = 6$ and $x = 8$ (circled) are equally far from the (true)
value ($\Delta I = 2$), but the lower ($x = 8$) point has a $1/\sigma^2$
weight almost twice that of the higher ($x = 6$) point; this contributes
to a lower \chisquare{} value for a model with a modest downward
deviation from the true model (blue dashed curve); an upward-deviating
model (red dashed curve) will be an even worse fit. \textit{Right
panel}: Error bars now show uncertainties according the the
\chisquaremodel{} approach ($\sigma = \sqrt{m}$) for the
upward-deviating model (red error bars) and the downward-deviating model
(blue error bars). The $x = 6$ and 8 points now have almost equal
weights, but the slightly larger error bars in the case of the
upward-deviating model help produce a lower \chisquare{} for that model.
The \chisquare{} values for the three models (fit to all the data
points) are shown in the upper-right part of each panel.
\label{fig:bias-demo}}

\end{figure*}

Figure~\ref{fig:bias-demo} provides a simplified example of how
the two forms of bias operate, using a Gaussian + constant background
model and a small set of Poisson ``data'' points. The original (true)
model is shown by the solid gray line, while the dashed red and blue
lines show potential models offset above and below the correct model,
respectively. The calculated \chisquaredata{} (left) and
\chisquaremodel{} (right) values for the offset models are also
indicated, showing that the downward-offset model has the
lowestchisquaredata{} of the three, while the upward-offset model has
the lowest \chisquaremodel{} value.

In both cases, the bias is strongest when the mean counts are low,
and so for S\'ersic fits affects the outer, low-surface-brightness part
of galaxy. In order to accommodate the downward bias of \chisquaredata{}
fits, S\'ersic models with lower $n$ and smaller $r_{e}$ (fainter and
steeper outer profiles) are preferred; the $I_{e}$ value increases in
compensation to ensure a reasonable fit in the high-surface-brightness
center of the galaxy, where the bias is minimal. The opposite trends
hold for \chisquaremodel{} fits.


%
%
%
%
%
%
%
%



\begin{figure*}
\begin{center}
\includegraphics[scale=0.88]{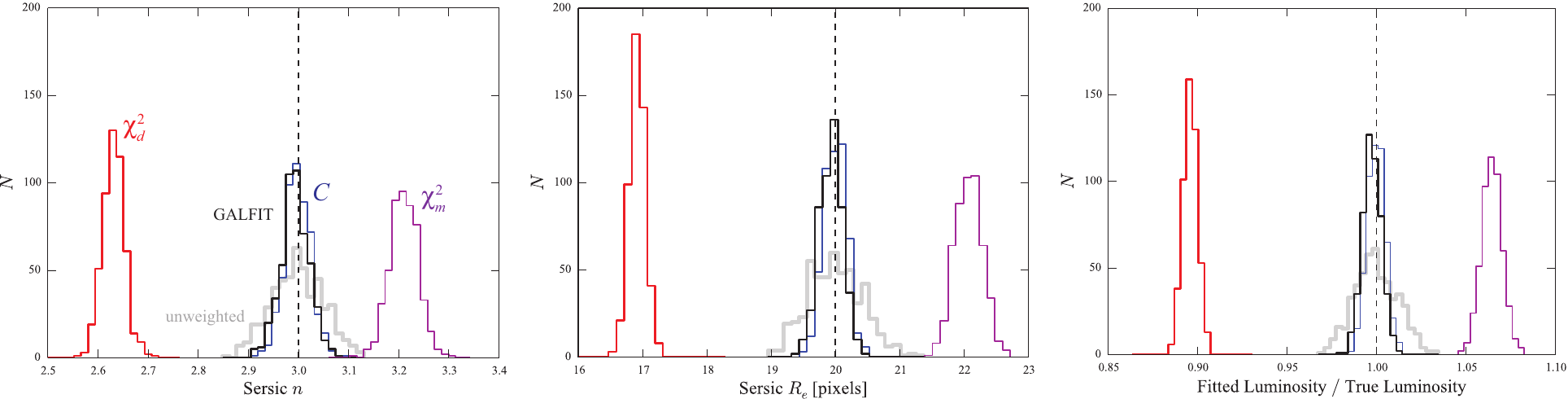}
\end{center}

\caption{Distribution of best-fit S\'ersic parameters $n$ and $r_{e}$
and total luminosity (relative to the true luminosity) from fits to
low-S/N model images; vertical dashed lines show original model values.
As in Figures~\ref{fig:sersic-histograms} and \ref{fig:lum-histograms},
data-based \chisquaredata{} fits (red histograms) underestimate $n$,
$r_{e}$, and luminosity, while model-based \chisquaremodel{} fits
(magenta) overestimate them; the Poisson MLE fits ($C$, blue) are
unbiased. \textit{Unweighted} \chisquare{} fits (thick, light gray
histograms) are unbiased but less accurate. Finally, fits using
\textsc{galfit} and its default $\sigma$ estimation (black) are almost
identical to the Poisson MLE \imfit{} results. Slight differences
in \chisquaredata, \chisquaremodel, and $C$ histograms with respect to 
those in the ``low-S/N'' panels of Figures~\ref{fig:sersic-histograms} and \ref{fig:lum-histograms}
are due to using uniform histogram bins within each panel.
\label{fig:sigma1-galfit-histograms}}

\end{figure*}

\subsection{Biases (or Lack Thereof) in Other Image-Fitting Software} 

%

It is important to note that the \chisquare{} biases  illustrated
above apply to specific cases of estimating Gaussian errors from the
data or model values on a pixel-by-pixel basis. They do \textit{not}
necessarily apply when an external error map is used in the fitting,
unless the error map is itself primarily based on the individual pixel values.
Error maps generated in other ways may have little or no effective bias.

For example, the default behavior of \textsc{galfit} is to compute an error map
from the image by estimating a global background RMS value from the
image (after rejecting a fraction of high and low pixel values),
combined with per-pixel $\sigma$ values from a background-subtracted,
\textit{smoothed} version of the data. This means that \textsc{galift}'s \chisquare{} 
calculation is actually
\begin{equation}
\chi^{2} \; = \; \sum_{i = 1}^{N} \frac{ (d_{i} - m_{i})^{2} }{ d_{i}^{\prime} + \sigma_{\mathrm{RMS}}^{2} },
\end{equation}
where $\sigma_{\mathrm{RMS}}$ is a global value for the image and $d_{i}^{\prime}$
is the background-subtracted, smoothed version of the data; the underlying
rationale is that a constant background should have, in the Gaussian approximation,
a single $\sigma$ value (Chien Peng, private communication).

This approach has two advantages over the simpler \chisquaredata{}
method. First, smoothing the (background-subtracted) data before using
it as the basis for $\sigma$ estimation helps suppress the fluctuations
which give rise to the \chisquaredata{} bias, as demonstrated by
\citet{churazov96} for the case of 1D spectroscopic data. Second, the
\textsc{galfit} version of the \chisquare{} statistic is similar in some
respects to the so-called \textit{modified} Neyman's \chisquare{}:
\begin{equation}
\chi^{2}_{N,\mathrm{mod}} \; = \; \sum_{i = 1}^{N} \frac{(d_{i} - m_{i})^{2}}{\mathrm{max}(d_{i}, 1)},
\end{equation}
where $d_{i}$ is the per-pixel data value.  In both cases, the effect of
a fixed lower bound to the error term ($\sigma_{\mathrm{RMS}}$ or 1) is
to transition from approximately Poisson weighting of pixels when
$d_{i}$ or $d_{i}^{\prime}$ is large to \textit{equal} weighting for
pixels when the object counts approach zero (this also removes the
problem of pixels with zero counts). This tends to weaken, though not 
eliminate, the bias in the low-count
regime \citep[e.g.,][]{mighell99,hauschild01}. A hint of this effect can
even be seen in the top panel of Figure~\ref{fig:sersic-histograms},
where the addition of a constant read-noise term to the $\sigma$
estimation reduces both the \chisquaredata{} and \chisquaremodel{}
biases.

Figure~\ref{fig:sigma1-galfit-histograms} shows the distribution
of $n$, $r_{e}$, and total luminosity for S\'ersic fits to the low-S/N
model images of Section~\ref{sec:model-images} for the \chisquaredata,
\chisquaremodel, and \pmlr{} ($C$) fits, along with fits to the same images using
\textsc{galfit} (version 3.0.5). The distributions from fits using
\textsc{galfit} (black histograms) are almost identical to those from
the \pmlr{} fits using \imfit{} (blue histograms). A very slight
bias in the \chisquaredata{} sense -- i.e., underestimation of S\'ersic
$n$, $r_{e}$, and luminosity -- can still be seen in the \textsc{galfit}
results, but this is marginal and in any case much smaller than the
dispersion of the fits. Similarly, some evidence for the same
biases can be seen in the \textsc{galfit} simulations of \citet[][their
Fig.~5]{haussler07}, \citet[][their Fig.~6]{hoyos11}, and \citet[][their
Figs.~4 and 5]{davari14}, but these deviations are only visible at
the very lowest S/N levels and are tiny compared to the overall scatter of
best-fit values.

The alert reader may have noticed that the discussion of how the
\textsc{galfit} approach reduces bias in the fitted parameters implies
that \textit{unweighted} least-squares fits should be unbiased. So would
it be better to forgo various $\sigma$ estimation schemes entirely, and
treat all pixels equally? Figure~\ref{fig:sigma1-galfit-histograms}
shows that \chisquare{} fits to the same (simple S\'ersic) model images
are indeed unbiased when all pixels are weighted (thick gray
histograms). But the drawback of completely unweighted fitting is clear
in the significantly larger dispersion of fitted results: unweighted
fits are less \textit{accurate} than either the Poisson MLE or \textsc{galfit}
approaches.





\section{Summary}\label{sec:summary} 

I have described a new open-source program, \imfit, intended for
modeling images of galaxies or other astronomical objects. Key features
include speed, a flexible user interface, multiple options for handling
the fitting process, and the ability to easily add new 2D image
functions for modeling galaxy components.

Images are modeled as the sum of one or more 2D image functions, which
can be grouped into multiple sets of functions, each set sharing a
common location within the image. Available image functions include standard
2D functions used to model galaxies and other objects -- e.g., Gaussian,
Moffat, exponential, and S\'ersic profiles with elliptical isophotes --
as well as broken exponentials, analytic edge-on disks, Core-S\'ersic
profiles, and symmetric (and asymmetric) rings with Gaussian radial
profiles. In addition, several sample ``3D'' functions compute line-of-sight
integrations through 3D luminosity-density models, such as an
axisymmetric disk with a radial exponential profile and a vertical
$\mathrm{sech}^{2/n}$ profile. Optional convolution with a
PSF is accomplished via Fast Fourier Transforms, using
a user-supplied \textsc{fits} image for the PSF.

Image fitting can be done by minimization of the standard \chisquare{}
statistic, using either the image data to estimate the per-pixel
variances (\chisquaredata) or the computed model values
(\chisquaremodel), or by using user-supplied variance or error maps.
Fitting can \textit{also} be done using Poisson-based
maximum-likelihood estimators (Poisson MLE), which are especially
appropriate for cases of images with low counts per pixel and low or
zero read noise. This includes both the traditional Cash statistic $C$
frequently used in X-ray analysis and an equivalent likelihood-ratio
statistic (\pmlr) which can be used with the fastest
(Levenberg-Marquardt) minimization algorithm and can also function as a
goodness-of-fit estimator. Other minimization algorithms include the
Nelder-Mead simplex method and Differential Evolution. Confidence
intervals for fitted parameters can be estimated by the
Levenberg-Marquardt algorithm from its internal covariance matrix; they
can also be estimated (with any of the minimization algorithms) by
bootstrap resampling. The full distribution of parameter values from
bootstrap resampling can also be saved to a file for later analysis.


A comparison of fits to artificial images of a simple S\'ersic-function
galaxy  demonstrates how the \chisquare-bias discussed by
\citet{humphrey09} manifests itself when fitting images: fits which
minimize \chisquaredata{} result in values of the S\'ersic parameters
$n$ and $r_{e}$ (as well as the total luminosity) which are biased low
and values of $I_{e}$ which are biased high, while fits which minimize
\chisquaremodel{} produce smaller biases in the opposite directions; as
predicted, these biases decrease, but do not vanish, when the background
and source intensity levels increase. Fits using Poisson MLE
statistics yield essentially \textit{unbiased} parameter values; this is
true even when Gaussian read noise is present. S\'ersic fits to images
of real elliptical galaxies with varying exposure times or background
levels show evidence for the same pattern of biased parameter values
when minimizing \chisquaredata{} or \chisquaremodel. This suggests
that the fitting of galaxy images with \imfit{} should generally use
Poisson MLE minimization instead of \chisquare{} minimization
whenever possible, especially when the background level is less than
$\sim 100$ photoelectrons/pixel.

Precompiled binaries, documentation, and full source code (released under the
GNU Public License) are available at the following
web site:\\ 
\href{http://www.mpe.mpg.de/~erwin/code/imfit/}{http://www.mpe.mpg.de/\~{}erwin/code/imfit/}.

\acknowledgements 

Various useful comments and suggestions have come from Maximilian
Fabricius, Martin K{\"u}mmel, and Roberto Saglia, and thanks are also
due to Michael Opitsch and Michael Williams for being (partly unwitting)
beta testers. Further bug reports, suggestions, requests, and fixes from
Giulia Savorgnan, Guillermo Barro, Sergio Pascual, and (especially)
Andr{\'e} Luiz de Amorim are gratefully acknowledged. I also thank
the referee, Chien Peng, for a very careful reading and pertinent
questions which considerably improved this paper. This work was partly
supported by the Deutsche Forschungsgemeinschaft through Priority
Programme 1177, ``Witnesses of Cosmic History: Formation and evolution
of galaxies, black holes, and their environment.''

This work is based in part on observations made with the
\textit{Spitzer} Space Telescope, obtained from the NASA/IPAC Infrared
Science Archive, both of which are operated by the Jet Propulsion
Laboratory, California Institute of Technology under a contract with the
National Aeronautics and Space Administration. This paper also makes use of
data obtained from the Isaac Newton Group Archive which is maintained as
part of the CASU Astronomical Data Centre at the Institute of Astronomy,
Cambridge.

Funding for the creation and distribution of the SDSS
Archive has been provided by the Alfred P. Sloan Foundation, the
Participating Institutions, the National Aeronautics and Space
Administration, the National Science Foundation, the U.S. Department of
Energy, the Japanese Monbukagakusho, and the Max Planck Society. The
SDSS Web site is \texttt{http://www.sdss.org/}.

The SDSS is managed by the Astrophysical Research Consortium (ARC) for
the Participating Institutions.  The Participating Institutions are
The University of Chicago, Fermilab, the Institute for Advanced Study,
the Japan Participation Group, The Johns Hopkins University, the
Korean Scientist Group, Los Alamos National Laboratory, the
Max-Planck-Institute for Astronomy (MPIA), the Max-Planck-Institute
for Astrophysics (MPA), New Mexico State University, University of
Pittsburgh, University of Portsmouth, Princeton University, the United
States Naval Observatory, and the University of Washington.

\appendix

\section{Comparing Levenberg-Marquardt and Bootstrap Estimates of Parameter Uncertainties for Image Fits} 

\subsection{Parameter Estimates for Fits to PGC~35772}

Table~\ref{tab:haggis} lists the best-fit parameters for three
progressively more complex models of the spiral galaxy PGC~35772 (see
Section~\ref{sec:haggis-fit} and Figure~\ref{fig:haggis}), along with
both the Levenberg-Marquardt (L-M) uncertainties and the uncertainties
derived from 500 rounds of bootstrap resampling (the latter listed in
parentheses after the L-M uncertainties). A comparison of the two types
of uncertainty estimates suggests they are similar in size for the
simplest model (fitting the galaxy with just a S\'ersic component), with
mean and median values of $\sigma_{\rm bootstrap} / \sigma_{\rm LM} =
1.38$ and 0.81, respectively. However, the bootstrap uncertainties are
typically about half the size of the L-M uncertainties for the more
complex models: the mean and median values for the uncertainty ratios
are 0.48 and 0.51, respectively, for the S\'ersic + Exponential model
and 0.61 and 0.41 for the S\'ersic + GaussianRing + Exponential model.


%
%

\subsection{Parameter Estimates for Multiple Exposures of Elliptical Galaxies}

In Section~\ref{ref:biases-real} I compared different \chisquare{} fits
with Poisson MLE fits for several elliptical galaxies. In this section,
I compare L-M and bootstrap parameter error estimates for two of the same elliptical
galaxies (plus a third observed under similar conditions), always using
Poisson MLE fits in order to avoid \chisquare{} bias effects.
Specifically, I compare best-fit parameters from S\'ersic fits to
multiple images of the same galaxy, in two ways.

\begin{figure*}
\begin{center}
\includegraphics[scale=1.05]{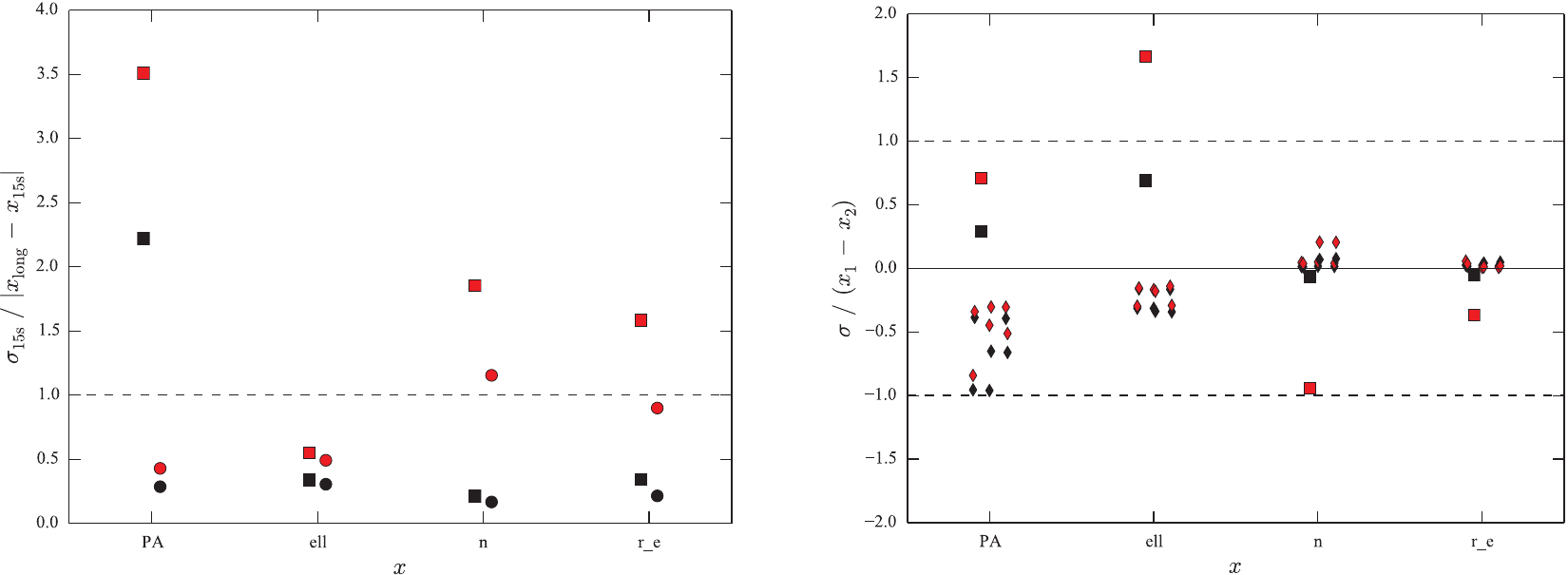}
\end{center}

\caption{Comparison of different estimates of parameter
uncertainties for S\'ersic fits to low- and medium-S/N images of
elliptical galaxies. \textit{Left:} Estimated parameter uncertainties
$\sigma_{15 {\rm s}}$ from fits to 15s images of NGC~3379 ($V$-band,
INT-PFCU; squares) and NGC~4697 ($r$-band, INT-WFC; circles), divided by
the difference in fitted parameter values between those fits ($x_{15
{\rm s}}$) and fits to longer exposures ($x_{\rm long}$ = mean of two
40s exposures for NGC~3379, single 60s exposure for NGC~4697). Black =
using $\sigma$ from Levenberg-Marquardt covariance-matrix; red = using
$\sigma$ from bootstrap resampling analysis. The general trend is for
the $\sigma$ values to underestimate the deviances; the
L-M estimates are always somewhat worse in this sense.
\textit{Right:} Estimated uncertainties from fits to two 40s $V$-band
images of NGC~3379 (squares) and three 15s $R$-band images of NGC~3377
(INT-WFC; diamonds), divided by the difference in parameter values
between the individual fits to the same galaxy (e.g., $\sigma_{n} /
(n_{2} - n_{1})$); colors as for the left panel. For NGC~3379, the
estimated errors from fits to the two images are very similar and only
average values are used; for NGC~3377, all combinations of errors and
differences from pairs of images are used.
\label{fig:uncertainty-ratios}}

\end{figure*}

First, I compare best-fit parameter values $x$ (e.g., S\'ersic index
$n$) from fits to short (15s) exposures with values from fits to longer
($2 \times 40$s or $1 \times 60$s) exposures with the same telescope +
filter system on the same night. I do this by comparing differences in
parameter values $\Delta x = x_{\rm long} - x_{\rm short}$ with the
error estimates for the same parameter $\sigma_{x}$ from the short
exposures (left panel of Figure~\ref{fig:uncertainty-ratios}). This can
be thought of as a crude answer to the question: how well do the
error estimates describe the uncertainty of parameters from
short-exposure fits relative to more ``correct'' parameters obtained
from higher S/N data? (I do not compare $I_{e}$ values because these can
vary due to changes in the transparency between exposures; similarly, I
do not compare values for the pixel coordinates of the galaxy center
because these depend on the telescope pointing and are intrinsically
variable.)

To first order, if the uncertainty estimates were reasonably accurate we
should expect $\sim 68$\% of the points to be found at $\sigma_{15 \mathrm{s}} /
|\Delta x| < 1$  and $\sim 32$\% to be at $ \sigma_{15 \mathrm{s}} / |\Delta x| >
1$. As the left panel of the figure shows, neither approach is ideal,
but the bootstrap-$\sigma$ estimates are somewhat better: 50\% of those
are $> 1$, while this is true for only 1/8 of the deviations if the L-M
$\sigma$ estimates are used.

The second approach, seen in the right-hand panel of
Figure~\ref{fig:uncertainty-ratios}, is to compare how well the
differences in parameter values obtained from similar samples (i.e.,
multiple images of the same galaxy with the \textit{same} exposure time)
compare with the error estimates. This is, in a limited sense, a test of
the nominal frequentist meaning of confidence intervals: how often do
repeated measurements fall within the specified error bounds? In this
case, I am comparing parameters from fits to the two 40s $V$-band
exposures of NGC~3379 (squares) and also parameters from fits to three
15s $R$-band exposures of the lower-luminosity elliptical galaxy
NGC~3377 (diamonds), also from the INT-WFC. Again, we should expect
$\sim 68$\% of the points to lie within $\pm 1$ if the error estimates
are accurate; as the figure shows, essentially \textit{all} the
bootstrap and L-M estimates lie inside this range and so tend
to be too small, particularly for $r_{e}$. The bootstrap estimates do a
better job in the case of NGC~3379 and a worse job in the case of
NGC~3377.

\subsection{Summary}

The implication of the preceding subsections is that the
L-M and bootstrap estimates of parameter errors are very
roughly consistent with each other, though there is some evidence that
the latter tend to become smaller than the former as the fitted models
become more complex (i.e., more components). In general,
both the L-M and bootstrap estimates should probably be
considered \textit{underestimates} of the true parameter uncertainties,
something already established for L-M estimates from tests of other
image fitting programs \citep[e.g.,][]{haussler07}.

\smallskip

\end{document}